\documentclass[manuscript]{emulateapj}
\usepackage{natbib}
\usepackage{float}
\usepackage{times}
\usepackage{amsmath}
\usepackage{amsfonts}
\usepackage{url}

\usepackage[usenames,dvipsnames]{color}
\usepackage{MnSymbol}
\usepackage{multirow}
\usepackage[normalem]{ulem}

\newcommand{\mstar}{${\cal M}_\star$}
\newcommand{\lmstar}{$\log{\cal M}_\star$}
\newcommand{\fnuc}{${\cal F}_{\rm nuc}$}
\newcommand{\mdens}{$\Sigma_{{\rm eff},{\cal M}_\star}$}

\definecolor{purple}{rgb}{0.4,0.0,1.} 
\definecolor{lightgreen}{rgb}{0.67,0.87,0.0} 
\definecolor{dodgerblue}{rgb}{0.12, 0.56, 1.0}

\slugcomment{To be submitted to The Astrophysical Journal}

\shorttitle{NGFS: II.~The Central Dwarf Galaxy Population}
\shortauthors{Eigenthaler et al.}

\begin{document}

\title{The Next Generation Fornax Survey (NGFS): II.~The Central Dwarf Galaxy Population}

\author{Paul~Eigenthaler$^{1,14}$\footnote{CAS-CONICYT Postdoctoral Fellow}, Thomas H.~Puzia$^{1}$, Matthew  A.~Taylor$^{2}$, Yasna~Ordenes-Brice\~no$^{1,6}$, Roberto~P.~Mu\~noz$^{1}$, Karen X.~Ribbeck$^{1}$, Karla~Alamo-Mart\'inez$^{1,}$\footnote{FONDECYT Postdoctoral Fellow}, Hongxin Zhang$^{1,}$\footnote{FONDECYT Postdoctoral Fellow}, Sim\'on~\'Angel$^{1}$, Massimo Capaccioli$^{4}$, Patrick C{\^o}t{\'e}$^{5}$, Laura Ferrarese$^{5}$, Gaspar Galaz$^{1}$, Eva K.\ Grebel$^{6}$, Maren Hempel$^{1}$, Michael Hilker$^{7}$, Ariane~Lan\c{c}on$^{8}$, Steffen Mieske$^{3}$, Bryan Miller$^{9}$, Maurizio Paolillo$^{10}$, Mathieu Powalka$^{8}$, Tom Richtler$^{11}$, Joel Roediger$^{5}$, Yu Rong$^{1,14,15}$, Ruben S\'anchez-Janssen$^{12}$, Chelsea Spengler$^{13}$}
\affil{
$^{1}$Instituto de Astrofísica, Pontificia Universidad Cat\'olica de Chile, Av.~Vicu\~na Mackenna 4860, 7820436 Macul, Santiago, Chile\\
$^{2}$Gemini Observatory, Northern Operations Center, 670 North A'ohoku Place, Hilo, HI 96720, USA\\
$^{3}$European Southern Observatory, 3107 Alonso de C\'ordova, Vitacura, Santiago\\
$^{4}$INAF-Osservatorio Astronomico di Capodimonte, Salita Moiariello 16, 80131, Naples, Italy\\
$^{5}$NRC Herzberg Astronomy and Astrophysics, 5071 West Saanich Road, Victoria, BC V9E 2E7, Canada\\
$^{6}$Astronomisches Rechen-Institut, Zentrum f\"ur Astronomie der Universit\"at Heidelberg, M\"onchhofstr.~12-14, D-69120 Heidelberg, Germany\\
$^{7}$European Southern Observatory, Karl-Schwarzchild-Str. 2, D-85748 Garching, Germany\\
$^{8}$Observatoire astronomique de Strasbourg, Universit\'e de Strasbourg, CNRS, UMR 7550, 11 rue de l'Universite, F-67000 Strasbourg, France\\
$^{9}$Gemini Observatory, South Operations Center, Casilla 603, La Serena, Chile\\
$^{10}$Department of Physics, University of Naples Federico II, C.U. Monte Sant'Angelo, via Cinthia, 80126, Naples, Italy\\
$^{11}$Departamento de Astronom\'ia, Universidad de Concepci\'on, Casilla 160-C, Concepci\'on, Chile\\
$^{12}$STFC UK Astronomy Technology Centre, Royal Observatory, Blackford Hill, Edinburgh, EH9 3HJ, UK\\
$^{13}$Department of Physics and Astronomy, University of Victoria, Victoria, BC V8P 5C2, Canada\\
$^{14}$Chinese Academy of Sciences South America Center for Astronomy and China-Chile Joint Center for Astronomy, Camino El Observatorio 1515, Las Condes, Santiago, Chile\\
$^{15}$National Astronomical Observatories, Chinese Academy of Sciences, 20A Datun Road, Chaoyang District, Beijing 100012, China
}
\email{paul.eigenthaler@icloud.com}

\begin{abstract}
We present a photometric study of  the dwarf galaxy population in the core region ($\lesssim\!r_{\rm vir}/4$) of  the Fornax galaxy cluster based on deep $u'g'i'$ photometry from the {\it  Next Generation Fornax Cluster Survey}.~All imaging
data were obtained  with the Dark Energy Camera mounted  on the 4-meter Blanco telescope  at the Cerro-Tololo Interamerican Observatory.  We identify 258 dwarf galaxy candidates  with luminosities $-17\!\lesssim\!M_{g'}\!\lesssim\!-8$\,mag,
corresponding to typical stellar masses of $9.5\gtrsim \log{\cal M}_{\star}/M_\odot \gtrsim 5.5$, reaching $\sim\!3$\,mag  deeper in point-source luminosity and $\sim\!4$\,mag deeper in surface-brightness sensitivity compared to the classic
Fornax Cluster Catalog. Morphological analysis  shows that the dwarf galaxy surface-brightness profiles are  well represented by single-component S\'ersic models with average S\'ersic  indices of $\langle n\rangle_{u',g',i'}=(0.78-0.83) \pm
0.02$, and  average effective radii  of $\langle r_e\rangle_{u',g',i'}\!=(0.67-0.70)  \pm 0.02$ kpc.  Color-magnitude relations indicate  a flattening  of the galaxy  red sequence at  faint galaxy luminosities,  similar to the  one recently
discovered in the Virgo cluster. ~A comparison with  population synthesis models and the galaxy mass-metallicity relation reveals that the average faint dwarf galaxy is likely  older than $\sim\!5$\,Gyr.~We study galaxy scaling relations
between stellar mass, effective radius, and  stellar mass surface density over a stellar mass range covering six orders  of magnitude. We find that over the sampled stellar mass range several  distinct mechanisms of galaxy mass assembly can
be identified: {\it i)} dwarf galaxies assemble mass inside the half-mass radius up to \lmstar~$\!\approx\!8.0$,  {\it ii)} isometric mass assembly in the range $8.0\lesssim \log{\cal M}_{\star}/M_\odot \lesssim10.5$, and {\it iii)} massive
galaxies assemble stellar mass predominantly in their halos at \lmstar~$\!\approx\!10.5$ and above.

\end{abstract}


\section{Introduction}
Representing the current standard  theory of structure formation in the universe,  the Lambda Cold Dark Matter ($\Lambda$CDM)  model is the simplest one that is  in general agreement with observations, e.g.\ the  large-scale structure distribution of
galaxies and their scaling relations. Although the $\Lambda$CDM paradigm is consistent with many of these phenomena at  large scales, tensions with observations arise at mass scales of $\lesssim10^9\,M_\odot$, where $\Lambda$CDM predicts many more DM
dominated dwarf galaxy  haloes than are actually observed  around giant galaxies. This discrepancy between  the numbers of predicted and  observed dwarfs is generally referred  to as the \emph{missing satellites problem}  \citep{moore,klypin} and its
origin is  still unclear.~Either the predictions  of $\Lambda$CDM are  not reliable -- possibly  arising from the spatial  resolution limits and  the treatment of baryonic  physics processes --  or many faint dwarf  galaxies have simply not  yet been
discovered. It was already  pointed out early on that various baryonic effects  like the impact of reionization indeed play  a significant role (Bullock et al. 2001).  Modern $\Lambda$CDM simulations that do account for a  variety of baryonic effects
including local  and global reionization,  star formation  and feedback, tidal  effects (e.g.\ Guo  et al.  2011; Brooks et  al. 2013; Sawala  et al.  2015, 2016) typically  achieve rather good  agreement with  the observed (dwarf)  galaxy luminosity
function.~Another interpretation is that the large majority of low-mass DM haloes have been very inefficient at star formation.

It  is now well-established that dwarf galaxies are predominantly found in galaxy group  and cluster environments which, respectively, tend
to host dwarfs of differing morphological types. For example, while dwarf ellipticals (dEs) dominate the cluster galaxy  population, star-forming dwarf irregulars (dIrrs) are most commonly found in the field or the cluster outskirts \citep{sandage84,
bin85, bin88}, suggestive of a link between dwarf galaxy evolution and the environment they reside in  \citep[e.g.][]{zhang12, mistani16, vdw17, read17}. This morphology-density or morphology-distance relation for dwarf galaxies appears to be largely
driven by tidal and ram pressure effects (Grebel, Gallagher, \& Harbeck 2003).

Dwarf galaxies have been detected throughout the Local Volume and beyond, see e.g.\ \citet{bin91,cot97,kar98,kar99,kar00,chi09,mul15} and  many, many more.~However, detections for the faintest dwarfs, i.e.\ dwarf spheroidal galaxies, have so-far
been mainly  limited to those within  the Local Group  (LG) due to their  low surface-brightness \citep[e.g.][]{mcconnachie12}.~Thus,  identifying and studying these  faintest dwarf galaxy systems  in nearby galaxy  clusters and groups is  crucial to
constrain $\Lambda$CDM cosmology. %\textcolor{red}{\sout{and address the missing satellite problem in a larger context.}}
%In order to quantify the  discrepancy, the faint-end slope of the galaxy luminosity function (LF) -- described by a \citet{schechter} function of the form
%$\phi ({M_B})\,dM \propto {10^{0.4({M^*} -  M)(\alpha + 1)}}dM$ -- provides a diagnostic tool,  as the faint-end slope $\alpha$ can be directly  compared with $\alpha\approx-2$, predicted by $\Lambda$CDM for the  mass spectrum of cosmological DM
%haloes \citep{moore, jenkins}.
Previous observations seem to confirm the  missing satellites problem when comparing the observed galaxy numbers %\textcolor{red}{\sout{by hinting  at comparatively flat faint-end slopes with  respect to the}} 
with the $\Lambda$CDM halo  mass function in environments covering a  range of galaxy
densities \citep[e.g.][]{pritchet, trenthamtully, fer16,  tay17}, including rich galaxy clusters. Potentially easing some  of this tension is the discovery in  recent years of ultra-faint dwarf (UFD) and dwarf  spheroidal (dSph) satellites throughout
the Local Group  \citep{willman, bel06, zucker06a, bel07,  zucker07, mcconnachie09, bel10, mcconnachie12,  bel14, bec15, drl15, kop15,  lae15, hom16}, and rich  dwarf galaxy systems around  nearby giant galaxies and  clusters \citep[][]{kar07, crn16,
mul15, mun15, mul16, ord16, san16}.

The newly discovered  UFDs appear to be  an extension of dSphs to  lower luminosities \citep[e.g.][]{mun15}, being  much fainter ($M_{V}\!\gtrsim\!-8\,$mag) and smaller  ($r_{\rm eff}\!\lesssim\!300\,$pc) than classical  dSphs.~UFDs have luminosities
comparable to  globular clusters  (GCs), which  are much  more compact ($r_e\!\lesssim\!10\,$pc).~Classical  GCs typically  have $M/L_V\!\approx\!2$  \citep[e.g.][]{mcl05, vandenven, baumgardt,  str11, tay15}  whereas in  contrast, UFD  kinematics reveal
$M/L_{V}\gtrsim100$,  indicative of  DM-dominated  systems \citep[e.g.][]{kleyna,  simongeha}.  ~In addition  to  the rich  populations  of  low-surface brightness  (LSB)  dwarf galaxies  in  the Local  Volume,  there have  also  been discoveries  of
\emph{ultra-diffuse} galaxies  (UDGs), first  found and described  by \citet{sandage84,imp88,fer88,bot91}.  More recent  discoveries of this  galaxy class  have been identified  in various  galaxy aggregates  like the Coma  and Virgo  galaxy clusters
\citep{vandokkum, koda, mihos}, the  Pisces-Perseus supercluster \citep{mar16}, a galaxy group \citep{mer16}, in  the galaxy cluster Abell 2744 \citep{jans17} and  Abell 168 \citep{rom17a}, and even outside of groups  and clusters \citep{rom17b}. The
low stellar masses ($\sim\!6\!\times\!10^{7}\,M_{\odot}$) and large radii ($1.5\!-\!4.6$\,kpc) of UDGs result in very  low surface brightness values in the range $\mu_{0,V}\approx\!26\!-\!28.5$\,mag\,arcsec$^{-2}$, making them challenging to detect. The
 existence of this mysterious new galaxy class in mostly dense galaxy cluster environments prompts  the obvious question of whether there might be similar populations in other galaxy aggregates, or whether they
are reserved for rich galaxy clusters only.

In the present work we attempt to address these questions  by investigating the properties of the low surface-brightness dwarf galaxy population in the inner region of the Fornax galaxy  cluster, one of the most nearby southern galaxy clusters, using
data obtained  as part of the  {\it Next Generation  Fornax Survey} (NGFS). The  survey constitutes of  deep, wide-field, multi-passband $u'g'i'$  observations taken with the  Dark Energy Camera  \citep[DECam;][]{fla15} mounted on the  4-meter Blanco
telescope at Cerro Tololo Inter-American  Observatory (CTIO). Given its proximity, Fornax is a goldmine  for studying the formation and evolution of galaxies and  other stellar systems, such as GCs and dwarf galaxies,  in a galaxy cluster environment
\citep{mun15,iod16,dab16,wittmann16}.~Compared to its northern  counterpart, the Virgo cluster, Fornax has twice the central galaxy density, half the velocity dispersion, and accordingly a distinctly
lower mass \citep[$7\pm2\times10^{13}\,M_{\odot}$;][]{ferguson89, schuberth}.  Furthermore, the core of Fornax is  dynamically more evolved \citep{churazov} and its early-type  (E/S0) galaxy fraction is significantly larger  ($\sim\!50\%$) than that of
Virgo ($\sim35\%$), considering member galaxies brighter than $M_B=-16$\,mag not  classified as dwarfs in the FCC \citep{ferguson89} and VCC \citep{bin85} catalogues. Noting the above, the Fornax cluster  is an excellent target to study faint baryonic sub-structures
in one  of the most nearby cluster environments. Throughout this work we utilize a  distance modulus of $31.51\!\pm\!0.03$ mag for Fornax, corresponding to a distance of  $\sim\!20$\,Mpc \citep{bla09}. Derived magnitudes refer to
the AB system.

\section{Observations}
\begin{figure}[t]
\centering
\includegraphics[width=\columnwidth]{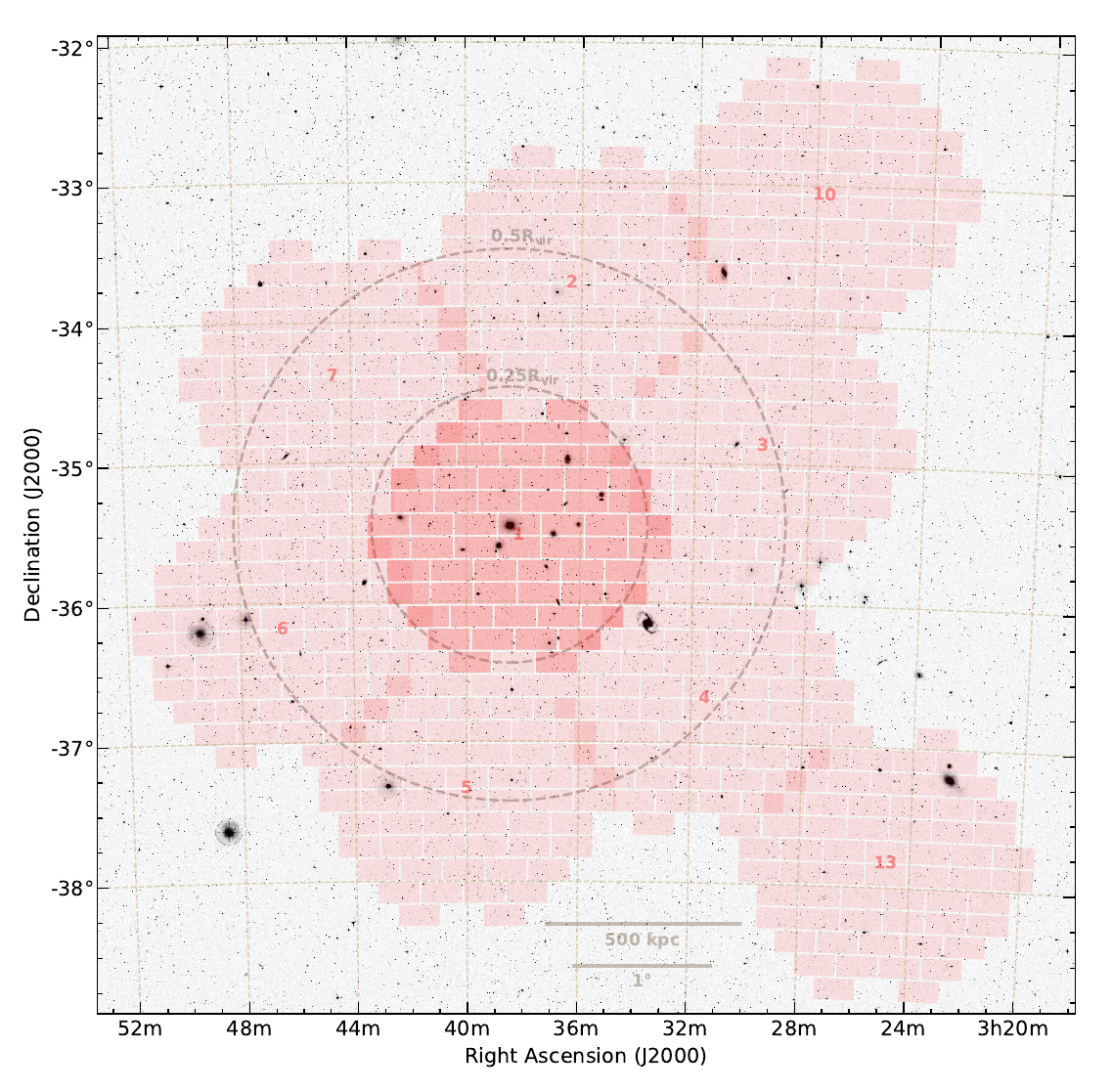}
\caption{Overview of the  NGFS mosaic footprint consisting  of DECam tiles $1\!-\!7,  10$, and $13$, with the central tile investigated in this work, being  highlighted.~The dashed circles  indicate a quarter and a half viral radius, 
i.e.~$0.25$ and $0.5\,R_{\rm vir}$.~The underlying figure is taken from DSS1 obtained via the Online Digitized Sky Surveys server at the ESO Archive.~The displayed field is $7^{\rm o}\!\times7^{\rm o}$.}
\label{ngfsmosaic}
\end{figure}

\begin{figure*}[ht]
\centering
\includegraphics[width=0.79\linewidth]{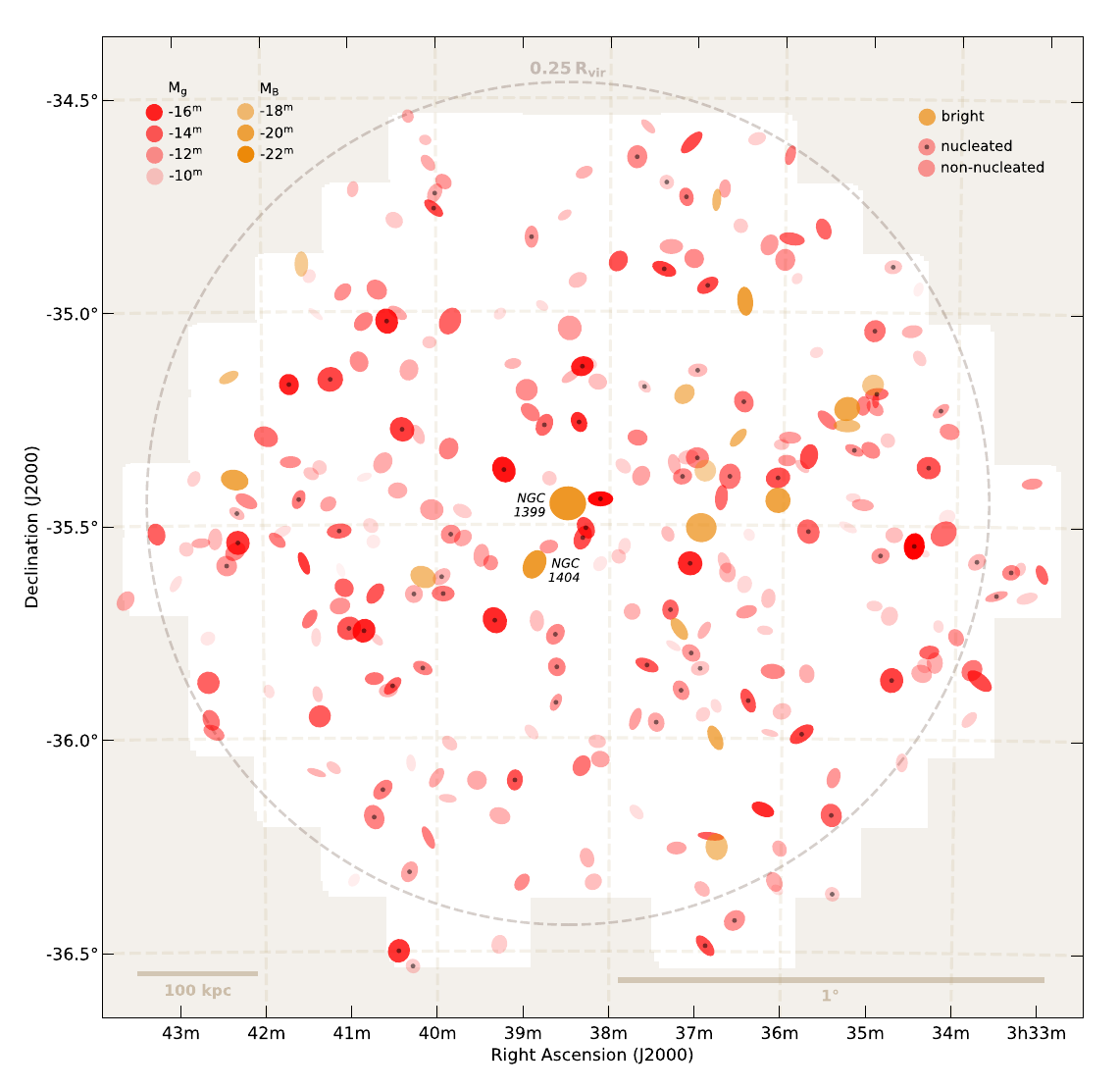}
\caption{Spatial footprint  of the central NGFS  tile centered  on NGC~1399 and analyzed in this  paper (shown in white).~The  distribution of the dwarf  galaxy population
corresponds to the one listed in Table\,\ref{tab:dwarfphot}.  Both bright galaxies ($M_B\leq -16$; orange) and dwarfs (red) are  indicated, and  the two brightest galaxies are labeled with their NGC numbers.  Ellipticities and position angles of the
symbols are scaled by the corresponding {\sc galfit} model values.~The dashed circle indicates $0.25R_{\rm vir}$ of  the Fornax cluster ($R_{\rm vir}\!=\!1.4$\,Mpc), as determined by \cite{dri01}.}
\label{spatial}
\end{figure*}

Our imaging is conducted  as part of the NGFS for which we provide  an abbreviated summary here to put  the present contribution into
context. The NGFS  is an  ongoing survey  of the  Fornax galaxy cluster  core in  the optical  $u'$, $g'$,  and $i'$ and  the near-infrared  (NIR) $J$  and $K_s$ filters,  targeting the  inner $30$\,deg$^2$,  corresponding to  $\sim\!1$\,Mpc in
galactocentric radius centered on  the cD galaxy NGC\,1399.~Here we focus  on the NGFS optical survey observations,  which consists of nine contiguous DECam tiles,  each aiming at a minimum point-source detection limit  of $u'=26.5$, $g'=26.1$, and
$i'=25.3$\,mag at  $S/N=5$ over the  PSF area. With the  goal of maximizing  sensitivity to LSB  structures in the imaging,  we employ the  Elixir-LSB dithering technique  developed for the complementary  {\it Next Generation  Virgo Survey}
\citep[NGVS, see][for details]{fer12}, and  use raw images processed by the DECam Community  Pipeline \citep[CP; v.2.5.0][]{val14} to produce fully calibrated image  stacks. The CP processes the raw DECam images  for basic calibration steps
(e.g.\ bias correction, flat  fielding, image crosstalk correction) and we use the  {\sc Astromatic}\footnote{\url{http://www.astromatic.net/software}} software suite \citep[{\sc Source Extractor}, hereafter {\sc  SE}, v.2.19.5; {\sc SCAMP}, v.2.0.1;
{\sc SWARP}, v.2.38.0;][]{ber96, ber02, ber06}  to perform the astrometric and photometric calibrations based  on reference stars from the 2MASS Point Source Catalog \citep{skr06} and SDSS stripe 82 standard star  frames, respectively, to build our image
stacks. We then apply our custom background subtraction strategy based on an iterative masking and sky modeling procedure.   We cross-validated our $u'g'i'$ photometry by running PSF photometry in the surveyed area and comparing the resulting magnitudes with the $U,B,V$ and $I$ photometry of \cite{kim13} for globular clusters in the same field.~In order to adequately compare the two photometries we utilized the empirical transformation equations from \cite{jordi06} and found good agreement within the uncertainties.

In a  previous contribution \citep{mun15},  we presented a preliminary  $i'$-band based analysis  of a rich  population of dwarf galaxies  in the central  $\sim\!3$\,deg$^2$ tile of  the NGFS footprint.~Here we  build upon these  results by
including the full  $u'g'i'$ color information for the central dwarf galaxy population with an average seeing  of 1.7\arcsec\  in $u'$, 1.3\arcsec\  in $g'$, and  1.1\arcsec\ in
$i'$. In order to evaluate the detection threshold for extended sources in $u'$, $g'$,  and $i'$, we directly measured the pixel statistics on the sky in the individual passbands, resulting in 1-$\sigma$ surface-brightness 
limits of $\mu_{u'}=28.04$, $\mu_{g'}=29.06$, and $\mu_{i'}=28.15$.~Figure~\ref{spatial} illustrates the  footprint of the central NGFS tile. 
 
\section{Analysis}
\subsection{Detection of LSB dwarf galaxies}
To detect LSB  dwarf galaxy candidates we first constructed  RGB images from the observed $u'g'i'$  frames to take advantage of  the total flux captured in each  passband while preserving color information.~We then  visually inspected these
frames looking explicitly for  diffuse LSB galaxies.~The RGB images allow us to  easily identify dwarf galaxy candidates by considering their  colors, sizes, and overall morphologies (e.g.\ comparatively flat  surface brightness profile and
nucleation). We also included LSB galaxies having SF knots and/or blue colours, not specifically selecting against these  sources. While automated algorithms can prove very effective in detecting faint galaxies (NGVS paper, Ferrarese et al, 2017,
ApJ, in press), for the purpose of  this work, we present a by-eye classification done independently by several people. ~The  advantage of this method over automated algorithms like {\sc SE} is obvious. For example,  {\sc SE} is only able to analyze
one frame--passband combination  at a time and  often fails to detect  extended LSB sources due  to contamination by foreground stars,  while visual inspection allows  for the identification of  galaxy candidates based on  all aforementioned criteria
simultaneously.~We find that galaxy color is particularly effective in  this regard since the colors of cluster dwarfs are expected to follow a cluster red sequence as seen  in many galaxy aggregates \citep{gladdersyee, roe17}, with likely origins in
the galaxy  mass-metallicity relation.  Combining this criterion  with the  diffuse morphologies arising  from shallow  dwarf surface-brightness  profiles results in  a straight-forward  detection strategy for  dwarf galaxy  candidates.~Despite these
advantages, the  disadvantage of visual inspection is that it precludes us from quantifying a selection function to verify sample completeness and that it has the the potential to introduce human bias.~Hence, to avoid any personal detection biases, we (PE, THP, YO, MAT,  KAM, KXR) individually investigated the RGB images independently, yielding up to six separate dwarf galaxy candidate catalogues for each tile. Tile 1 has five independent candidate catalogues.

\subsection{Quality flags}
To estimate the reliability of our detections we match the  catalogues using {\sc TOPCAT} \citep{topcat} to quantify in how many catalogues a given source was found.~We then assign  quality flags to each dwarf galaxy candidate such that one
found in  all five catalogues is  assigned a flag  $\mathcal{A}$, in four catalogues  a flag $\mathcal{B}$, and  so on.~The matching  is done systematically so  that we first  match all flag $\mathcal{A}$  objects, and remove them  from the
corresponding catalogues.~We continue by  searching for flag $\mathcal{B}$ sources, and continue until  all remaining sources are flag $\mathcal{E}$, i.e.\ untrustworthy  candidates only identified by a single team member.  In order to find
all matches  for a given flag  $k$, all possible  $\binom{5}{k}$ catalogue combinations  have to be matched,  leading to lists  of 145 flag $\mathcal{A}$,  59 flag $\mathcal{B}$,  54 flag $\mathcal{C}$,  29 flag $\mathcal{D}$, and  136 flag
$\mathcal{E}$ sources.

\begin{figure}[t!]
\centering
\includegraphics[width=1.05\columnwidth]{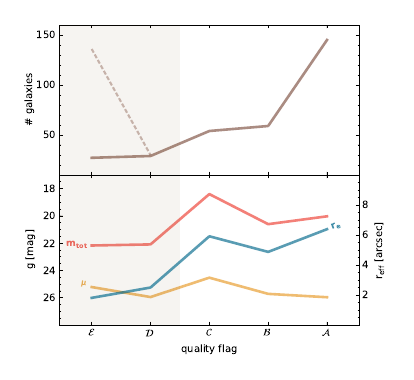}
\caption{Dwarf galaxy candidate detection flags.  \emph{Upper panel:} Distribution of detected dwarf galaxy candidates as a  function of the quality flags. For flags $\mathcal{E}$ the solid line indicates the average number of sources uniquely identified by one person, while the dashed line shows all accumulated flag $\mathcal{E}$ objects. \emph{Lower panel:} Total galaxy magnitudes  (red), surface brightness (yellow), and
effective radii (blue) as a function of quality flags. Sources with flags $\mathcal{ABC}$ are considered dwarf galaxy candidate detections. See text for details.}
\label{flags}
\end{figure}

Figure\,\ref{flags} shows the  distribution of quality flags, which illustrates the  potential human detection bias given the large number  of accumulated flag $\mathcal{E}$ sources present in the  merged candidate catalogue.~Conversely, those
sources with multiple independent  identifications, i.e.\ flags $\mathcal{ABCD}$, show a strong trend  toward unanimous detections. Flags $\mathcal{E}$ illustrate the average number of sources uniquely identified by one person (solid line) as well as all accumulated sources (dashed line).

The lower panel in  Figure\,\ref{flags} shows the  correlation between detection flags  and median values  of galaxy apparent $g'$-band  magnitude, effective radius ($r_e$),  and surface
brightness ($\mu_{g'}$) for each flag. As expected,  apparent magnitude and $r_e$ show clear trends toward brighter/larger galaxies having flags $\mathcal{A/B}$, while  flags $\mathcal{D/E}$ correspond to fainter/smaller sources. No obvious
correlation is seen with $\mu_{g'}$.~We  note that these correlations should be taken with  caution, since we only measure structural parameters for  a small subsample of $\mathcal{DE}$ flags. Moving forward,  we consider only galaxies with
detection flags $\mathcal{ABC}$, i.e.\  at least three independent identifications to  be dwarf galaxy candidates, yielding a  final sample of 258 sources. Of these  candidates, 75 ($\sim\!29$\%) show evidence of  nucleation based on visual
inspection of the $u'$, $g'$, and $i'$-band images, while 183 galaxies ($\sim\!71$\%) show no nucleus.

\subsection{Comparison with existing catalogues}
We compare our  sample with galaxies flagged  as \emph{likely members} in the  Fornax Cluster Catalogue \citep[FCC][]{ferguson89}.~Out  of 340 FCC galaxies, 112  fall in our observed DECam  footprint.~We note that we  could not re-identify
FCC\,162, which is  reported to be located within the  halo of the bright elliptical NGC\,1379  (FCC\,161), leaving 111 recovered FCC galaxies.  Of these, 90 are fainter than  $M_B=-16$\,mag and, hence, can be considered  as dwarf galaxies
\citep{ferguson94, tammann94}. In addition, two galaxies with $M_B\lesssim-16$ (FCC\,136 and FCC\,202) have explicitly been classified as dwarfs  in the FCC catalogue, resulting in a total number of 92 FCC dwarfs in our observed footprint.
We also recover 45 galaxies from the \citet{mie07} catalogue that  have not been classified as FCC galaxies. Finally, we checked for \citet{mie07} dwarfs that are not in our  sample and conclude that, based on their morphologies and colors,
all are likely to be background galaxies. Summarizing, we find a sample of $258-92-45=121$ previously uncatalogued dwarf candidates. We note that this number is smaller than in our recent publication \citep[see][]{mun15}, since in the present work we only consider 
galaxies with quality flags $\mathcal{ABC}$ as dwarf identifications.

Figure\,\ref{spatial} shows a schematic  representation of the spatial distribution of the  detected dwarf galaxy candidates and known  bright galaxies in the central DECam tile of the NGFS footprint analyzed in  this work (tile 1 in
Fig.~\ref{ngfsmosaic}). The symbol  shape, alignment, size, and opacity  of each galaxy marked in  Figure~\ref{spatial} is scaled by the  corresponding $\epsilon$, PA, $r_e$, and  $M_i$, respectively, as
obtained from our {\sc galfit} models.  We point out that since precise 3D spatial information is  not known, $M_g$  may vary by up to $\sim\!0.16\,$mag due to the unknown  depth of the cluster\footnote{This applies only under the
assumption of spherical symmetry.}.

\begin{figure*}[t!]
\centering
\includegraphics[width=\textwidth]{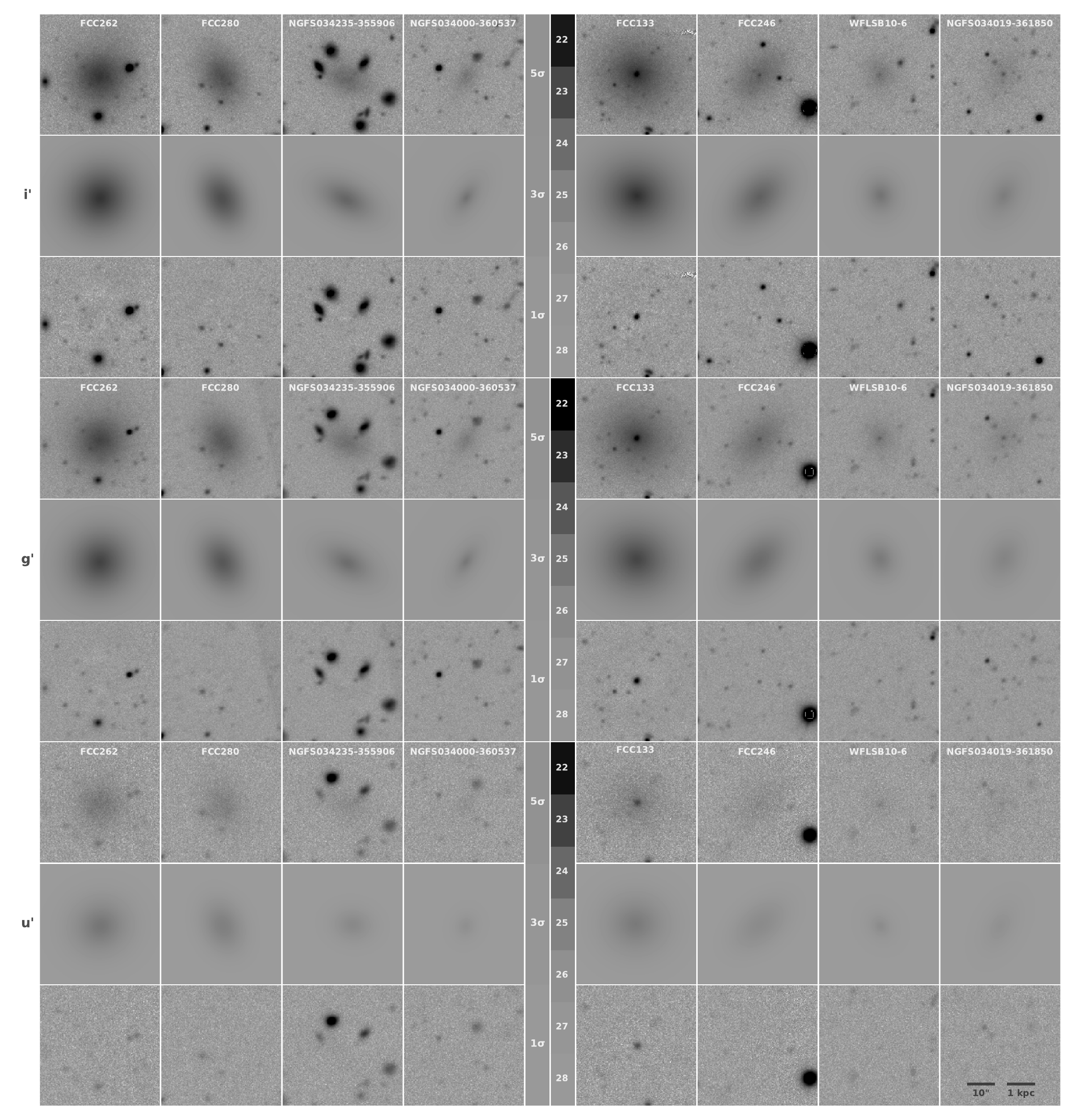}
\caption{Examples of the dwarf galaxy surface-brightness modeling with  {\sc galfit}. Four non-nucleated (left) and four nucleated (right) dwarfs with successively decreasing (left to  right) effective surface brightness values are shown in u' (lower panels), g' (middle panels), and i' (upper panels) as
examples. We show the dwarf galaxy postage stamps, the corresponding 2D \citet{ser68} models and residual images (galaxy--model) for every passband.
The greyscales show values  of constant surface brightness ranging  from $\mu=22$ to $28$  mag arcsec$^{-2}$ in steps of  1 mag arcsec$^{-2}$ as well as $1\sigma$, $3\sigma$, and $5\sigma$ surface-brightness detection thresholds in all passbands.~Note  that only the spheroid component  is modelled for nucleated  dwarfs so that the nuclear  star cluster is visible  in the residual
images. We also point  out the clear variance in the nucleus-to-galaxy  luminosity ratio for these dwarfs. Dwarfs  are f.l.t.r.:\,FCC262, FCC280, NGFS034235-355906, NGFS034000-360537, FCC133, FCC246,  WFLSB10-6, NGFS034019-361850. The images
cover $0.72\arcmin$ ($\sim4.25$\,kpc) on a side.}
\label{galaxymodeling}
\end{figure*}

\subsection{Surface Brightness Profile Modeling}
We determine  the structural parameters  for the 258  dwarf candidates  in $u'$, $g'$,  and $i'$ filters  utilizing the software  package {\sc galfit}  \citep[v3.0.5;][]{pen10}. {\sc  galfit} is an  effective tool for  extracting structural
information of dwarf galaxies, allowing the  user to model the two-dimensional distribution of the diffuse starlight in  galaxies with numerous parametric functions.~Utilizing the full 2D information for a  given galaxy enables us to obtain
reliable fits to the surface-brightness distribution constrained by each non-masked image pixel.~Surface-brightness profiles of dwarf galaxies are commonly well-fit by a single-component \citet{ser68} model
\begin{equation}
\label{eq:sersic}
 I(r) =  {I_e}\exp \{  - {b_n}[{(r/{r_e})^{{1 \mathord{\left/{\vphantom {1 n}} \right.\kern-\nulldelimiterspace} n}}} - 1]\}
\end{equation}
which is parametrized by the effective radius $r_e$, the intensity  $I_e$ at $r_e$, and the shape index $n$ (also referred to as the S\'{e}rsic index) that defines the curvature of the  model. The parameter $b_n$ is linked to $n$ such that half of the total  light from the model is enclosed within $r_e$
\citep{cao93}. We choose this single component, ellipsoidal model to measure the global morphology of our dwarf galaxy candidates.

Our  iterative fitting procedure  is split into several steps. First we create postage  stamp images for each dwarf candidate in the  $u'$-, $g'$-, and $i'$-band frames and construct corresponding segmentation  maps using {\sc SE}. The
segmentation maps are used to create bad-pixel masks for each dwarf, masking any non-dwarf sources above a $2\sigma$ threshold. This initial step results in dwarf-only images that are used for subsequent model fits.

\subsubsection{Automatic and Refined Fitting}
In a first iteration we fit galaxies  assuming four different initial guesses for the S\'{e}rsic model parameters. By comparison  between the four resulting model fits for each galaxy we obtain a  qualitative estimate of the robustness of a
fit. In the best scenario, all initial  guesses converge to the same solution, but in most cases two or three  initial guesses converge to a single model, leaving one or two outliers. In the  worst cases, all initial guesses yield different
results.~To properly assess the reliability of the  fits, we compare the galaxy, model, and residual images for each dwarf  in detail. In most cases where multiple initial guesses converge to the same  result, visual inspection shows the fits to
be reliable.

In the second  iteration we consider galaxies for which  the automatic fitting procedure did  not converge to a common  solution due to their diffuse natures  and/or small extents. For these  galaxies --- the majority of our  sample --- the
models display a severe mismatch with respect to the surface-brightness distributions of the galaxies, or the residual images show signs of over-subtraction. We refine the fitting procedure for these galaxies as follows.

First, we estimate the  total galaxy luminosity by running {\sc SE}  on the galaxy postage stamp images and  use the resulting MAG\_AUTO values from the  corresponding segmentation as initial guesses for the  {\sc galfit} S\'{e}rsic models.
Keeping these values fixed reduces the number of free parameters available for the next galaxy model, stabilizing the fit  and resulting in more robust estimates of the remaining model parameters. We then use the new parameters and fix them
and allow {\sc  galfit} to recompute the galaxy  luminosity freely. In the final  step, the newly determined galaxy  magnitude was again fixed, and  we recompute the other parameters  resulting in model parameters that  are all consistently
derived by {\sc galfit}. In this way, all non-nucleated dwarfs are fit successfully, leaving little to no sign of the galaxies' light in the residual images (see Fig.\,\ref{galaxymodeling}).

\subsubsection{Fitting Nucleated Dwarfs}
We modify the  above strategy for dwarf candidates that  show evidence of nucleation in order to account  for the excess light at their  centers.~While in principle a two-component model  consisting of a \citet{king62} profile for  the nucleus and a
S\'ersic profile  for the spheroid  would be preferred,  the pixel-scale sizes  of the nuclei  preclude this approach.~Very  few stable solutions are  found for the  nuclei, and the  addition of model  parameters increases the  likelihood of
degenerate solutions, making this approach futile.

As a  solution to this problem,  for a given  nucleated candidate, we instead  mask the nucleus  and fit a  single S\'{e}rsic model to  the diffuse spheroid  iteratively. We first manually  create a preliminary  mask for the nucleus  in the
segmentation maps and use  it to fit the galaxy light without  the nuclear component. From the resultant residual  images (including the nucleus) we generate an improved  mask for the nucleus and any other  sources contaminating the diffuse
components of the dwarfs. We then repeat the fit using the improved mask, and iteratively improve the mask and residual  image. If the nucleus is not masked correctly, the S\'{e}rsic model will attempt to fit it partially, thus predicting a
too high $n$, and producing symmetric regions of over-  and under-subtraction in the residuals. Following this procedure, we find that three to four iterations result in  clear nucleus--spheroid separation free of symmetric residuals in the
circum-nuclear regions, and give stable solutions for the spheroid components of all nucleated dwarf candidates confirmed by visual inspection.

Figure\,\ref{galaxymodeling} shows models for eight (four non-nucleated and four nucleated) representative  dwarf galaxies with successively decreasing average surface-brightness in the range $24.0\lesssim\mu_{i'}/{\rm mag}\lesssim26.5$.~Comparison of the galaxies (top row),
models (middle row), and residuals (bottom row) clearly illustrates the robustness of our modeling approach. In total,  we derive structural properties for 246/258 (95\%) of the detected dwarf candidates in $i'$- and $g'$-bands, dropping to
144 (56\%) in the  $u'$-band. While the lower fraction of  modeled galaxies in the $u'$-band is  likely due to lower sensitivity and comparatively  low flux of galaxies in this  particular passband, the 12 missing galaxies  in the $i'$- and
$g'$-bands are either  too faint, or too strongly contaminated  by a bright nearby star.~We list  our photometric results in Table\,\ref{tab:dwarfphot}, including  galaxy coordinates, all available total galaxy  magnitudes derived from {\sc
galfit}, integrated galaxy colors, absolute $g'$-band magnitudes  ($M_{g'}$), and estimated total stellar masses (\mstar; see also \S\,\ref{sec:masses}).~We also list the cross-identifications  from the FCC and the \citet{mie07} catalogues.
Likewise, we list in Table\,\ref{tab:dwarfmorph} all available structural parameters derived by {\sc galfit} in this work, also indicating for each galaxy whether or not a nucleus is present.

\section{Results}

\subsection{Structural Parameters}
\begin{figure*}[t]
\centering
\includegraphics[width=\textwidth]{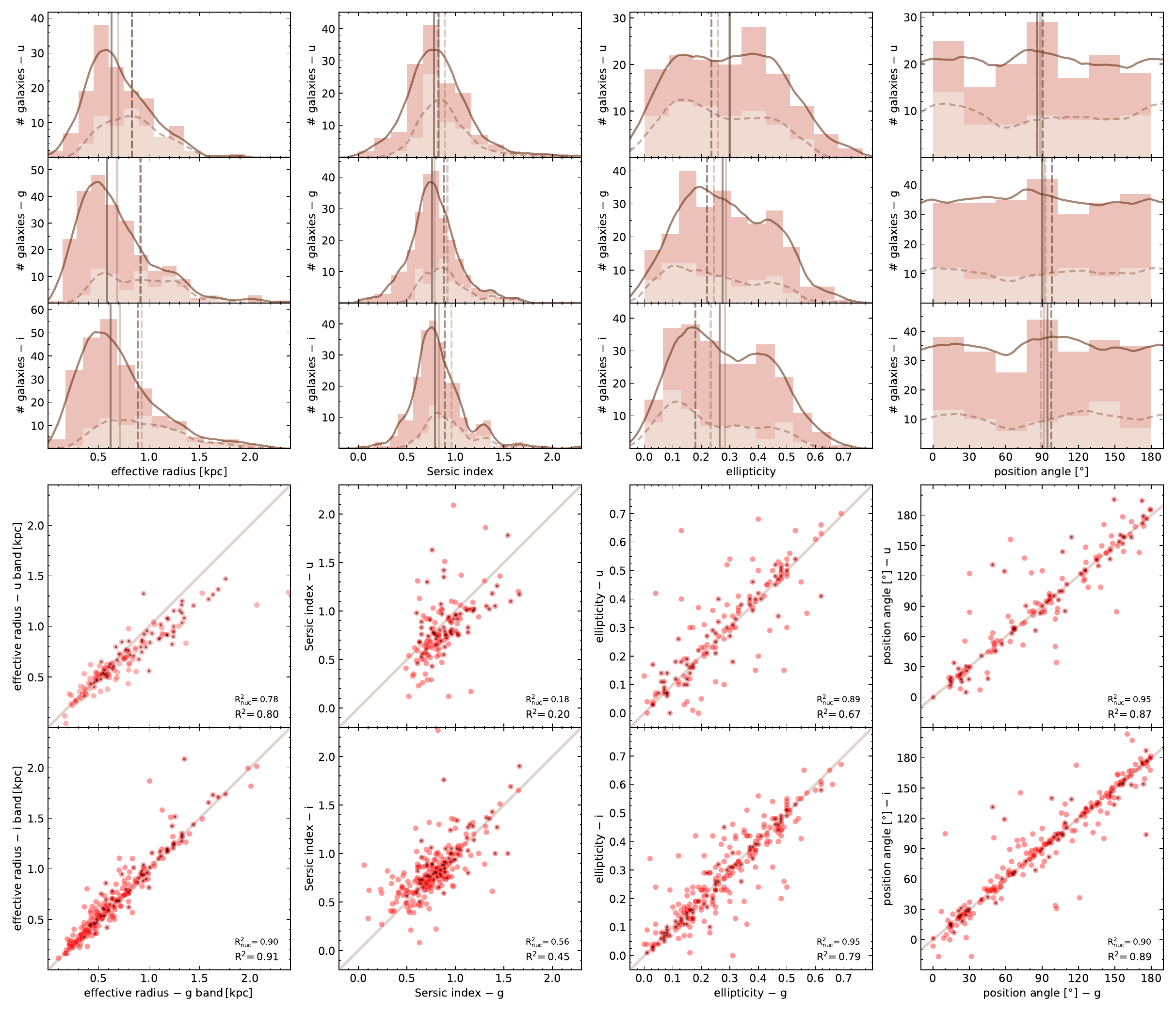}
\caption{Structural parameters derived from  {\sc galfit} in the $u'$, $g'$, and  $i'$ filters. {\it Upper panels}:\,the distributions for the  effective radius ($r_e$), S\'ersic profile shape parameter  ($n$), ellipticity ($\epsilon$), and
the position angle (PA)  in all three passbands. Lighter histograms refer to  nucleated dwarfs only. Smooth curves show the corresponding  Epanechnikov-kernel probability density estimates for the entire sample  (solid) and nucleated dwarfs
only (dashed). Vertical lines indicate the  mean (lighter) and median (darker) of the overall distributions for the  entire sample of dwarf galaxies (solid lines) and nucleated dwarf only  (dashed lines). {\it Lower panels}:\,the comparison
of $u'$- and $g'$-band parameters as  a function of the corresponding $i'$-band values. Straight lines show the unity  relations. Symbols with black dots refer to nucleated dwarfs. Correlation coefficients $R^2$  are shown in each panel for
both the overall sample and the nucleated dwarfs only.}
\label{structural}
\end{figure*}

We show in Figure~\ref{structural} the distribution functions of the  structural parameters derived by {\sc galfit} for the entire dwarf galaxy sample in the $u'$, $g'$  and $i'$-filters.~The upper panels show the distributions of effective
radius ($r_e$), S\'ersic profile shape parameter ($n$), ellipticity ($\epsilon$),  and the position angle (PA), split by filter with $u'$ in the top row followed by $g'$ and  $i'$ below.~The histogram bin-widths were chosen based on Knuth's
rule \citep{knuth}.~Alternative visualizations of the distributions are  shown by the solid relations, which are non-parametric {\it Epanechnikov}-kernel probability density estimates  (KDEs) for each parameter--filter combination. The lower
panels show consistency plots for each structural parameter, as derived from the $u'$, $g'$ and $i'$-filter.

While $r_e$  and $n$  show well-defined  peaks in their  distributions, the  $\epsilon$ and  PA values are  more broadly  distributed. We find  a characteristic  effective radius  consistent between the  passbands in  the range  $0.1\!\leq\! r_e/{\rm
kpc}\!\leq\!2.0$   with  a   similar   mean  size   of   $\langle  r_e\rangle_{g'}\!=\!0.68\!\pm\!0.02$\,kpc,   $\langle  r_e\rangle_{i'}\!=\!0.71\!\pm\!0.03$\,kpc, and       $\langle r_e\rangle_{u'}\!=\!0.69\!\pm\!0.03$\,kpc,  which is at first impression indicative of a lack of significant stellar population  gradients in the spheroid components.
~We find  in  all filters  similar  minimum  drop-offs in  $r_e$  and similar  $r_{e,{\rm max}}\approx2$\,kpc. 
However, the $r_e$ consistency plots in the bottom panels show that the $u'$-band half-light radii are systematically smaller than in the redder bands. This may be due to the lower surface-brightness depth of our $u'$-band data and/or stellar population gradients pointing towards bluer cores of dwarf galaxy spheroids.~Quantitatively evaluating the detailed causes of this intriguing result goes beyond the scope of this paper. We will, however, return to this aspect in a future paper on the stellar population content of the NGFS dwarf galaxies.
The distributions of S\'ersic indices $n$ closely follow Gaussian profiles in all passbands with  similar mean values of  $\langle  n\rangle_{u'}=0.83\pm0.03$, $\langle  n\rangle_{g'}=0.78\pm0.02$, and  $\langle
n\rangle_{i'}=0.83\pm0.02$. However,  we note that there may be  an observational bias giving rise  to the narrow distributions in $n$,  likely due to
selection effects since we explicitly looked for diffuse galaxies in our search for dwarf galaxy candidates.

We  find generally  more  platykurtic distributions  in  $\epsilon$  and PA  than  in $r_e$  and  $n$.~Our measurement  values  range between  $0.0\!\lesssim\!\epsilon\!\lesssim\!0.7$,  with  indications that  $\epsilon$  may be  bi-modally
distributed.~Despite the visual impression, detailed Gaussian mixture modeling (GMM) does not conclusively quantify whether a single-mode or a bimodal distribution is preferred.~We find a marginal tendency of increasing average ellipticity towards bluer passbands.

Interestingly, we do not find any highly elongated dwarfs,  with 97\% of the NGFS dwarfs exhibiting ellipticities $\epsilon < 0.55$. This is in very good agreement with the result from \cite{san16} for dwarfs in the core of the Virgo cluster, and with the fast rotators in the {\sc atlas3d} sample \citep{cap16}.
Finally, we see that position angles do not show any preferred alignment and populate the whole parameter range $0\degr \lesssim \rm{PA} \lesssim 180\degr$ with very good consistency between the $u'$, $g'$ and $i'$-filter.

The lower panels in Figure\,\ref{structural} compare the derived structural  parameters in all three passbands. Assuming the galaxies are homologous in all passbands, i.e.\ same $r_e$, $n$, $\epsilon$,  and PA, one would expect all galaxies to follow
unity, with the observed  scatter arising from the statistical uncertainties of  the measurements. However, this would also imply  that the dwarf candidates do not exhibit color  gradients, which is not necessarily true, as  pointed out above. Hence,
while the statistical measurement errors likely  contribute to the observed scatter, intrinsic differences between galaxy models --  e.g.\,color gradients -- may also play a role. In general, the $i'$  vs.\ $u'$-band comparison shows a systematically
larger scatter than  that of the $i'$  vs.\ $g'$-bands, in part likely  due to the shallower  $u'$-band observations. Detailed consideration of  the derived $R^2$ values  reveals that $r_e$ and PA  show the tightest correlations,  implying that these
structural parameters constrain the galaxy models most efficiently in either  of the filters. Meanwhile, the $n$ distribution shows a much larger deviation from unity, indicating that it is  more difficult for {\sc galfit} to find one unique solution
for the S\'ersic index in all passbands. In any case,  we note that the range in $n$ is relatively small, i.e.~83\% of the measured galaxies have shape parameters in  the range $0.5\!\lesssim\!n\!\lesssim\!1.5$, and variations on the order of $\Delta
n\simeq 0.1$ are expected.

We also highlight the nucleated dwarfs in the parameter distributions of the upper panels of Figure~\ref{structural}, which clearly  show that the average spheroid of a nucleated dwarf has a larger half-light radius and more spherical light
distribution, with a hint of a slightly more  exponential-type S\'ersic profile (i.e. $n$ closer to 1) than their average non-nucleated counterpart. There is mild indication that the spheroid components of nucleated dwarf galaxies are more spheroidal, i.e. show lower ellipticity, than their non-nucleated counterparts.
~ Except for effective radii, the nucleated dwarf samples show tighter correlations than the overall samples in the filter correlation plots. This is primarily due  to the fact that nucleated dwarfs are on average brighter than non-nucleated  dwarfs. We find that the nucleation fraction (\fnuc) is a strong function  of galaxy luminosity. \fnuc\ shows a strong
tendency towards  higher fractions in bins  containing the brightest  galaxies, and declines to  $\sim0\%$ when only  the faintest galaxies are  considered. Likewise, the cumulative  \fnuc\ shows a  similar trend, in that  \fnuc$=100\%$ for
exclusively bright galaxies, and falls to the final value of $\sim29\%$ for the overall dwarf galaxy sample. The same trend  of higher nucleation fractions in brighter dEs was found in the Virgo cluster by \cite{grant05} and \cite{lisker07}, indicating a possibly, generally valid trend. We furthermore note that, given the present seeing, low-luminosity nuclei that have low contrast with the host galaxy could remain unidentified and affect the derived nucleation fraction.

\subsection{Color-Magnitude Relations}
\label{sec:cmd}
\begin{figure*}[t!]
\centering
\includegraphics[width=\textwidth]{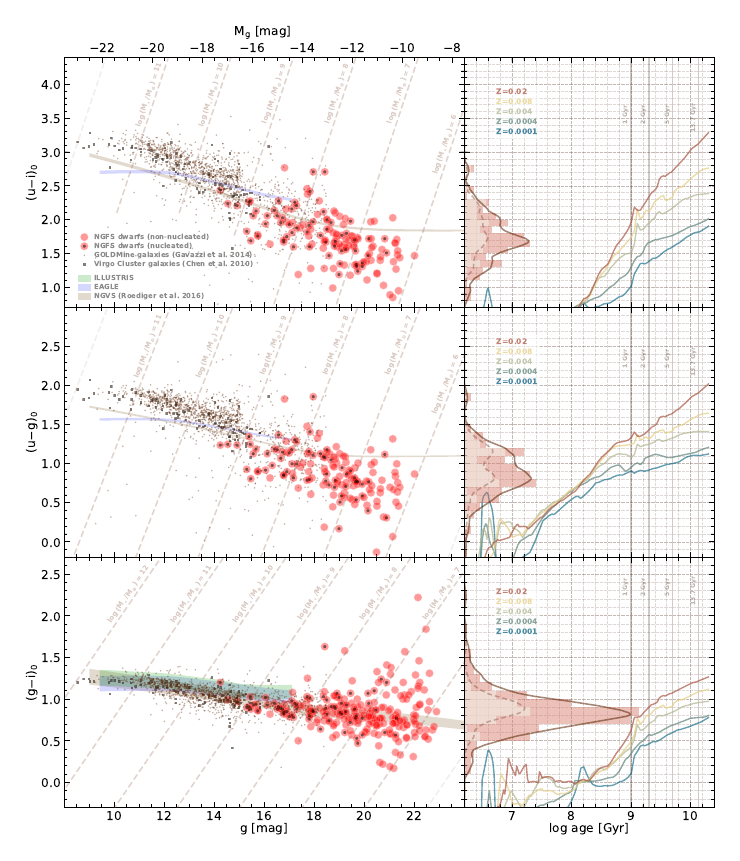}
\caption{Color-magnitude diagrams and simple  stellar population model predictions for stellar populations  in NGFS galaxies. \textit{Left panels:} Color-magnitude relations  based on $u'$, $g'$, and $i'$-band photometry.  
Dashed lines show \mstar\ estimates from \citet{bell03}, while solid lines show red sequence fits  from NGVS \citep{roe17}. \textit{Right panels:} Color distributions are shown along the ordinate of each panel for
both the overall (dark shading) and  nucleated dwarfs (light shading) samples. Also overplotted are \cite{bc03} model  predictions relating SSP ages and metallicities with the corresponding galaxy colors.  Vertical lines indicate ages at 1,
2, 5, and 13.7\,Gyr.}
\label{cmds}
\end{figure*}

Color-magnitude diagrams (CMDs) of  galaxies have long been used to infer the  mass assembly and star formation histories of galaxies  in large-area, deep surveys with ground and space-based  telescopes \citep[e.g.][]{bal04, bell04, bell12,
fab07,  tra16,  roe17}.~Multi-passband  CMDs  are  powerful  tools   to  obtain  constraints  of  luminosity-weighted  average  stellar  population  ages  and  metallicities,  as  well  as   stellar  masses  of  large  samples  of  galaxies
\citep[e.g.][]{chilingarian12}.

In the left panels  of Figure\,\ref{cmds} we show the $(g'\!-\!i')_0$,  $(u'\!-\!g')_0$, and $(u'\!-\!i')_0$ vs.\ $g'$ CMDs  for NGFS dwarf galaxies and compare  them with other galaxy samples and  theoretical predictions of color-magnitude
relations.~We account  for foreground extinction  toward Fornax by utilizing  average extinction
coefficients  ($A_{u'}$,  $A_{g'}$, and  $A_{i'}$)  derived  from several  bright  Fornax  galaxies based  on  the  re-calibrated extinction  maps  of  \citet{schlafly}, noting  that  the  extinction variance  across  the  observed field  is
$\sigma_{A_{u'}}\!\simeq\!0.008$,  $\sigma_{A_{g'}}\!\simeq\!0.006$,  and  $\sigma_{A_{i'}}\!\simeq\!0.003$\,mag.~Dashed  lines  show  the  iso-\mstar\  relations  based  on  the  \citet{bell03}  color-${\cal  M}/L$  conversions  (see  also
Sect.~\ref{sec:masses}).~Given the photometric  limits of the combined  dataset the sample reaches  down to stellar masses  $\sim\!10^6 M_\odot$, while the  optical $g'$ and $i'$-band  photometry reaches even lower  values.~All three panels
clearly illustrate the expected red sequence for early-type  cluster galaxies, where brighter systems exhibit redder colors than fainter ones.~This color-magnitude relation of  galaxies is commonly interpreted as a mass-metallicity relation
since the deeper potential wells  of more massive galaxies more easily retain metals produced throughout their star-formation history \citep[e.g.][]{kod97, pog01, tre04, sav05, mai08, bal08, mol13, seg16, ma16}.

We fit linear  relations to the CMDs of NGFS  dwarfs and derive shallow color gradients  in the sense of $\Delta(g'  - i')_0/\Delta g' = {\nabla _{(g'  - i')_0}} = -0.01$, $\Delta(u' - g')_0/\Delta g' = {\nabla _{(u'-  g')_0}} = -0.09$, and $\Delta(u' - i')_0/\Delta g' ={\nabla _{(u'  - i')_0}} =
-0.13$.

~We note that  the scatter present around  the red sequence likely  arises from effects intrinsic  to the constituent stellar  populations (e.g.~spreads in ages/metallicities)  and/or from photometric uncertainties.  To quantify the
scatter in  the observed CMDs  we computed the  variance of the  measured values with  respect to  the linear fits,  obtaining $\sigma_{(g' -  i')_0}\!=\!0.25$, $\sigma_{(u' -  g')_0}\!=\!0.24$, and $\sigma_{(u'  - i')_0}\!=\!0.27$\,mag.
~We  discuss below (see Sect.~\ref{sec:ssp}) the implications of  these color variances in
terms of stellar population properties.

\subsubsection{Comparison with other Environments}
In order to put our Fornax measurements in a  larger context, we overplot in Figure\,\ref{cmds} the red-sequence fits for Virgo cluster galaxies from the NGVS \citep[see][]{roe17}. This  dataset is comparable to our NGFS data in photometric
depth,    SED    sampling    and     homogeneity.~For    this,    we    convert    the    NGVS    photometry    to     the    standard    SDSS    system    utilizing    the    photometric    transformation     equations    on    the    CADC
website\footnote{\url{http://www.cadc-ccda.hia-iha.nrc-cnrc.gc.ca/en/megapipe/docs/filt.html}. We stress that that the transformation from the CFHT/MegaCam to the SDSS $u$-band is known to be biased, in the sense that the SDSS $u$-band covers substantially bluer wavelengths than the MegaCam u-band.~Thus, the corresponding transformation between the filter sets becomes an extrapolation rather than an interpolation.}.~The brown  shaded area illustrates  the uncertainty in  the photometric transformation arising  from multiple transformation  equations and filter throughput mismatch.~We also
plot earlier CMD relations for the  100 ACS Virgo Cluster Survey (ACSVCS) galaxies based on SDSS  measurements \citep{che10}.~In addition, we show measurements from the GOLDMine database  \citep{gav03, gav14} for various other environments,
including the Coma, Cancer, Hercules, A2197, and A2199 galaxy clusters, as well as the Local Supercluster.

With the  combined empirical dataset, we  note a relatively  sharp red cutoff for  galaxies at the  upper envelope of the  red sequence in  the $(g'\!-\!i')_0$ vs.~$g'$ CMD, which  continues from bright  galaxies with $10^{11}M_\odot$ down  to faint
luminosities of our NGFS dwarf  galaxy sample at $10^7 M_\odot$ and lower masses.~Galaxies  above this threshold are potentially heavily reddened or  possible background sources, and their offsets from the  red sequence toward redder colors
are even more obvious in the $(u'\!-\!g')_0$ and $(u'\!-\!i')_0$ CMDs, likely a result of internal dust absorption and/or redshift in combination with the wider SED sampling.

We note that the  NGVS red sequence fit lies systematically towards redder colors compared to the present  NGFS $(u'\!-\!g')_0$ and $(u'\!-\!i')_0$ data.~Despite the photometric filter transformation uncertainties, one might speculate whether this discrepancy is  due to a systematically
lower metallicity of the dwarf galaxy population in Fornax compared to Virgo.~But we defer further discussion on this interesting aspect until spectroscopic metallicities become available.

\subsubsection{Comparison with Numerical Model Predictions}
In order  to test  whether the  overall empirical datasets  shown here  are in  agreement with  current galaxy formation  models, we  overplot theoretical  color-magnitude red-sequence predictions  from the  Illustris and  EAGLE simulations
(\citealt{vogelsberger,schaye15}).~Illustris consists of eight cosmological N-body hydrodynamic  simulations, each spanning a volume of $\sim\!10^{6}$ Mpc$^{3}$, differing in terms of  resolution, among other things, while EAGLE comprises
six simulations  similar to Illustris,  running a modified  SPH code.~In  order to ensure  a consistent comparison  between these simulations,  our dataset, and  the NGVS red-sequence  fits we  utilize the same  model relations as  shown in
\citet{roe17}, i.e.~adopting the same selection  criteria for obtaining the samples of Illustris and EAGLE galaxies.~The  galaxy populations from both numerical models cover a range of  ${\cal M}_\star\gtrsim 10^{8.5}M_\odot$ while our NGFS
dwarf sample  reaches masses down to  ${\cal M}_\star\gtrsim 10^{6}M_\odot$.~The green  and blue-shaded bands  illustrate the uncertainty in  the photometric transformation  from NGVS photometry to  the standard SDSS system.
Both the Illustris and EAGLE  simulations predict systematically redder colors than the NGFS  dataset and the NGVS fits for galaxies fainter than $M_{g'}\!\approx\!-18$ mag ($\sim\!10^{10}\,M_\odot$).~We observe that no model reproduces the  red sequence slope in Fornax' and Virgo's core \citep[see][]{roe17}, with model slopes being systematically shallower than observed.
We conclude that this discrepancy could be explained by shortcomings of  the galaxy formation models.~Unfortunately, given the small overlap in mass between these models and the NGFS  dataset, a statistically robust comparison between the red-sequence model
slopes and the present data is not feasible.

\subsubsection{Stellar Population Properties}
\label{sec:ssp}
The right-hand panels  of Figure~\ref{cmds} show an attempt  to constrain the (luminosity-weighted) ages of  the dwarf candidate sample by  comparing their broad-band colors with  the prediction of simple stellar population  (SSP) models of
\citet{bc03}\footnote{\url{http://www.bruzual.org/bc03/}}, hereafter BC03.~These models  provide color predictions for \citet{cha03}, \citet{kro01},  and \citet{sal55} stellar initial mass functions (IMFs)  using BaSeL \citep{basel}, STELIB
\citep{stelib}, and MILES  \citep{miles} stellar libraries.~We compare  the various SSP permutations for  $(g'\!-\!i')_0$, $(u'\!-\!g')_0$, and $(u'\!-\!i')_0$  colors, finding that changing IMFs  has minimal effect on  the predicted galaxy
colors (average of maximum deviations  in all colors and metallicities is of order  $\sim\!0.02$ mag), while switching stellar libraries results in  more significant, but still minor differences (average of  maximum deviations in all colors
and metallicities is $\sim\!0.10$ mag). See \citet{pow16} for more comparisons between model predictions.~In the following we utilize the BC03 models based on the MILES stellar library with a Chabrier IMF.

We further  explore the color distributions  with the histograms  and associated Epanechnikov-KDEs shown  along the ordinates  in the right  panels of Figure~\ref{cmds}, for  both the overall  (red shading) and nucleated  (light shading)
samples.~While the  overall population shows generally  symmetric color distributions about the  peaks, the nucleated subsample  exhibits a tendency towards redder average colors.~This feature likely
arises from the  mass-metallicity relation, in light of  the trend toward higher  nucleation fractions with increasing luminosity  shown in Figure~2 of \cite{mun15},  and is consistent with  similar findings for dwarf galaxies  in the Virgo
cluster \citep{grant05}.

We plot the BC03  SSP model predictions for metallicities in the range $0.0001\!\leq\!Z\!\leq\!0.02$ as a function of  age.~The models clearly demonstrate the well-known age-metallicity degeneracy
that complicates stellar population  studies based on broad-band optical colors.~Nonetheless, relating our  derived $(g'\!-\!i')_0$ colors with the BC03 tracks,  we see that even at the highest metallicity  ($Z\!=\!0.02$), the average dwarf
shows ages $\gtrsim\!1$\,Gyr.~Given the mass--metallicity relation for  galaxies, it is likely that the vast majority of the low-luminosity dwarf sample   have significantly lower metallicities.~\citet{kirby} measured spectroscopic
metallicities from  individual red-giant branch  stars in fifteen MW  dSphs, seven LG dIrrs,  and thirteen M31 dSphs,  finding that their metallicities  scale as $\left\langle {[{\rm{Fe}}/{\rm{H]}}}  \right\rangle\! =\! -1.69 +  0.30\log \left(
{{{\cal M}_\star}/{{10}^6}{{\rm{M}}_ \odot  }} \right)$.~Comparing this relation with  metallicities of more massive galaxies  from \cite{gal05}, \citeauthor{kirby}~concluded that this  relation is roughly continuous from  the least massive
systems at  \mstar$=10^{3.5}\,M_\odot$ to the most  massive giant ellipticals at  \mstar$=10^{12}\,M_\odot$.~Our most massive  dwarf galaxy has \lmstar~$=9.42$,  corresponding to a metallicity  of $[{\rm Fe}/{\rm H}]\!\simeq\!-0.66$,  if it
follows the \citet{kirby} relation.~Considering standard abundances from \cite{asplund},  this converts to a metallicity of $Z\!\simeq\!0.003$.~Hence, based on the BC03 model $(g'-i')_0$ model tracks,  one can expect ages of at least 5\,Gyr
for the average dwarf  in our NGFS sample. While showing the  expected low metallicities, we note that the  dwarf candidate sample exhibits a large spread  in metallicity if it is mostly uniformly  old, i.e.~older than $\sim\!10$\,Gyr.~This
might be explained by the host cluster environment that is likely to have experienced epochs of  significant chemical inhomogeneities and, hence, may have uniquely influenced the star formation histories of the individual galaxies.~However,
the  difficulty in  breaking  the age--metallicity  degeneracy  with $u'g'i'$  photometry  alone is  obvious.~Future  imaging campaigns  that sample  different  regions of  the  spectral energy  distribution  \citep[e.g.][]{mun14} and  deep
integrated-field spectroscopy \citep[e.g.][]{men16}  will be very valuable  in further constraining the stellar  population properties of these low-mass  NGFS galaxies to higher  accuracy.~We, thus, defer a more  detailed stellar population
analysis and a discussion of the build-up of the faint galaxy population \citep{bos14}, including near-infrared photometry information, to future work.

\subsection{Stellar masses}
\label{sec:masses}
\begin{figure}[t]
\centering
\includegraphics[width=\columnwidth]{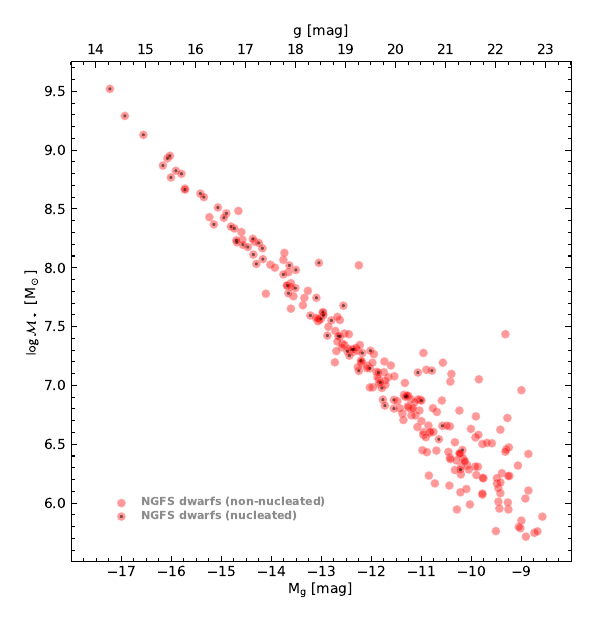}
\caption{The stellar mass--luminosity relation for our sample dwarf galaxies. Stellar masses are estimated using the prescription of \citet{bell03} utilizing the derived and dereddened $(g'\!-\!i')_0$, $(u'\!-\!g')_0$, and $(u'\!-\!i')_0$ colors. Black dots denote nucleated dwarfs.}
\label{stellarmass}
\end{figure}

In order to study the physical scaling relations of  the dwarf galaxy sample in relation with other stellar systems, we compute the stellar masses of our NGFS  dwarf candidates using the prescriptions from \citet{bell03}, which are good matches for the expected star formation histories of our NGFS spheroidal dwarf galaxies \citep{zha17}. We parametrize the
stellar mass-to-light ratios (\mstar$/L_\odot$) by galaxy colors using the relation

\begin{equation}
\label{eq:stellarmass}
\log {({{\cal M}_\star}/L)_ \odot } = {a_\lambda } + {b_\lambda } \times \rm{color}
\end{equation}
where the  coefficients $a_\lambda$ and $b_\lambda$  define the transformation  for different filter combinations,  i.e.~colors \citep[see][their Table\,7]{bell03}.~We compute  the $g'$ and  $i'$-band \mstar$/L_\odot$ from our  measured and
de-reddened $(g'\!-\!i')_0$,  $(u'\!-\!g')_0$, and $(u'\!-\!i')_0$  galaxy colors, yielding  up to six  \mstar\ estimates for  a given  galaxy.~To compute \mstar,  we derive galaxy  luminosities in $g'$  and $i'$ considering  absolute solar
magnitudes  measured by  C.\  Willmer\footnote{\url{http://mips.as.arizona.edu/\~{}cnaw/sun.html}}.~Finally, we  average  the  individual \mstar\  estimates  and adopt  error  bars as  the  standard  deviation of  the  sample of  individual
measurements.~We   obtain   stellar  masses   in   the   range  $5.5\lesssim   \log{\cal   M}_{\star}/M_\odot   \lesssim  9.4$.~Figure\,\ref{stellarmass} shows the comparison of the derived  stellar masses with the apparent and absolute $g'$-band magnitudes.~The corresponding \mstar\ values,  in the following always given in units
of $M_\odot$, are summarized in Table\,\ref{tab:dwarfphot}.

\begin{figure*}[t!]
\centering
\includegraphics[width=0.95\linewidth]{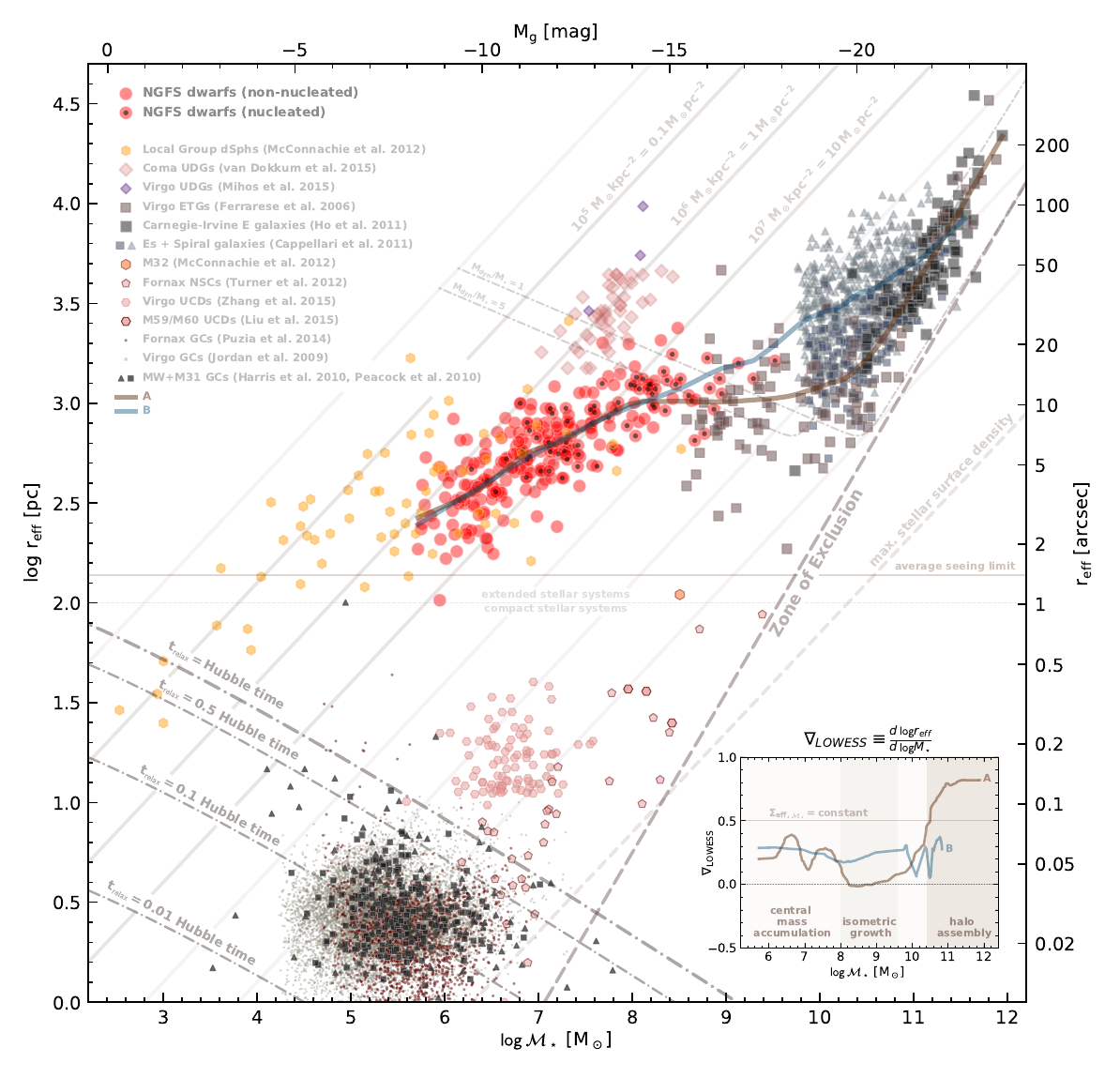}
\caption{Effective radius ($r_{\rm eff}$ in  units of parsec) vs. stellar mass (\mstar\ in  units of $M_\odot$) parameter space for NGFS galaxies  (red symbols) and various other stellar systems (see  legend) with lines of iso-\mdens$\equiv
{M_\star}/(2\pi r_e^2)$ indicated.~The top ordinate shows the  corresponding absolute $g'$-band luminosity scaled by the average relation from Figure~\ref{stellarmass}.~The parameter space is  split into extended and compact stellar systems
at $r_e\!=\!100$\,pc highlighted by  the horizontal dotted line. We also show the  average seeing limit of our observations measuring  $\sim1.4$ arcsec.~Curves of equal relaxation timescales for  $r_e-{\cal M}_\star$ parameter combinations
are indicated for  various fractions of the  Hubble time (thick dash-dotted curves).~The  Zone of Exclusion (thick  dashed line) illustrates the stellar  density limit beyond which virtually  no objects are found.~The line of maximum stellar surface density ($\Sigma_{\rm max}\approx10^{11.5}M_\odot$\,kpc$^{-2}$) observed for dense stellar systems in the nearby universe is also indicated \citep[see][]{hop10}.~Similar  relations for the
ATLAS3D datasets from \cite{cap16}  are plotted for two dynamical to stellar  mass ratios, ${\cal M}_{\rm dyn}/{\cal M}_\star=1$ and  $5$.~The brown curve indicates a LOWESS fit  to the NGFS dwarfs and other early-type  galaxy data from the
literature, which the blue curve is a LOWESS fit to the NGFS dwarfs and massive spiral galaxies only.~The inset plot  in the bottom right corner shows the gradients of these curves with three stellar mass zones dominated by different galaxy
mass accumulation processes (see text for details).}
\label{scalingrelations_reff}
\end{figure*}

\begin{figure*}[t!]
\centering
\includegraphics[width=0.95\linewidth]{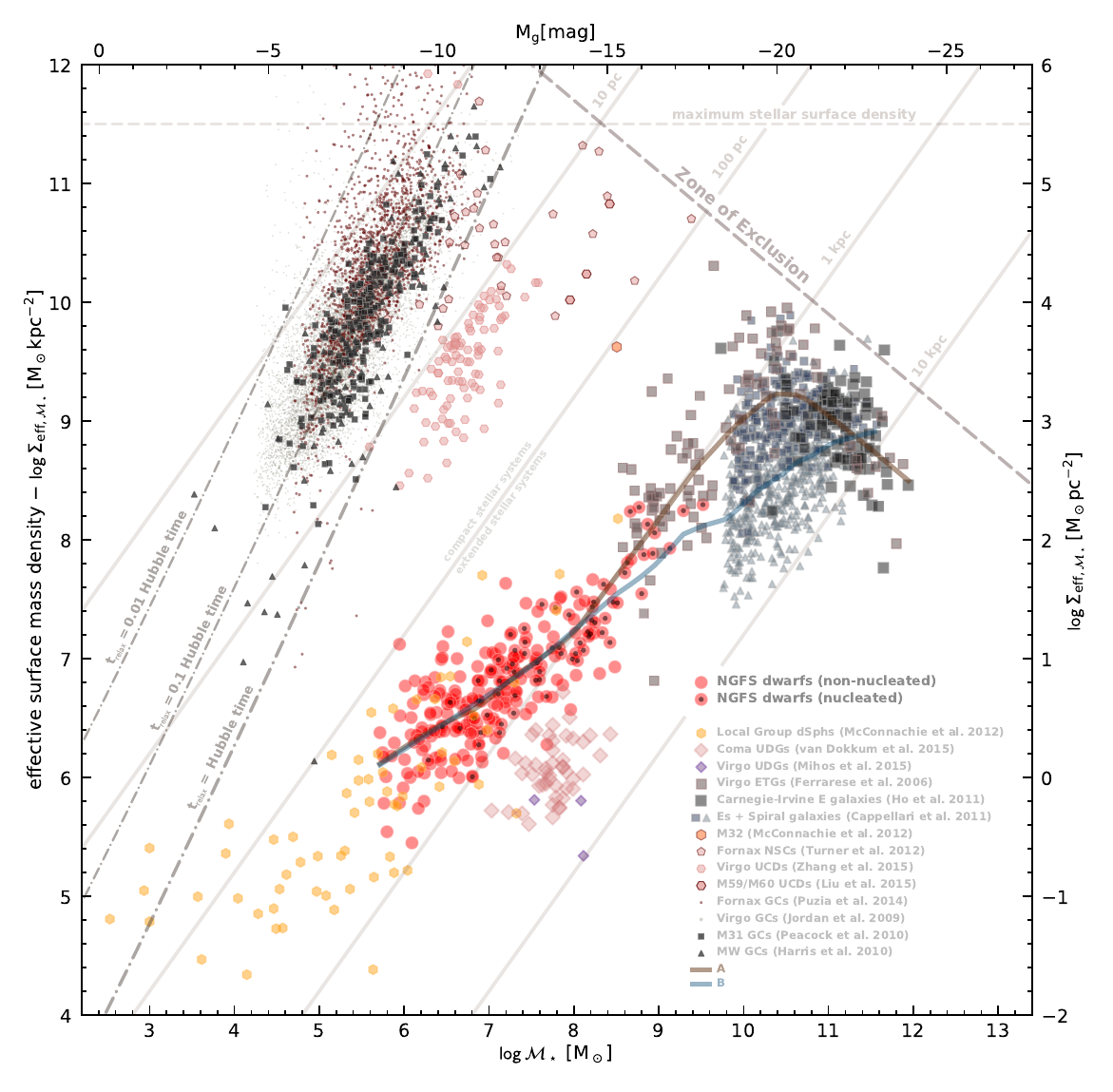}
\caption{Effective  surface mass  density (\mdens)  vs.~stellar  mass (\mstar)  parameter space  for  NGFS galaxies  (red  symbols) and  various other  stellar  systems (see  legend). Solid  diagonal  lines mark  constant effective  radius,
i.e.~iso-$r_{\rm  eff}$ lines.  Other curves  and lines  are  corresponding to  those in  Figure~\ref{scalingrelations_reff}.~The  top ordinate  shows the  corresponding absolute  $g'$-band  luminosity scaled  by the  average relation  from
Figure~\ref{stellarmass}, while the right abscissa gives the effective surface mass density in units of $M_\odot/{\rm pc}^2$.}
\label{scalingrelations_surfmd}
\end{figure*}

\section{Discussion}
\label{sec:discussion}
\subsection{Scaling Relations}
Elliptical galaxies  are known  to fall on  a so-called  fundamental plane (FP),  a tight  correlation between effective  radius, central  velocity dispersion, and  the average  effective surface brightness  within the  effective radius \citep[e.g.][]{faberjackson, faber87, dre87, djor87, bender92, bur97, gal06}.~A similar relation was found by \citet{tho11}, showing that dispersion-supported galaxies form a one-dimensional \emph{fundamental curve} in the mass-radius-luminosity space from ultra-faint dwarf spheroidals to giant cluster spheroids.~Since the FP shows very little residual  scatter it implies very similar mass-to-light ratios and suggests a uniform formation process for bright elliptical galaxies.~Projecting the FP onto the luminosity vs.~surface-brightness plane, \citet{kormendy85} had noted that there seems to be a dichotomy between the scaling relations of bright elliptical galaxies and dwarf galaxies.~This dichotomy can be interpreted as the result of different formation mechanisms for giant ellipticals and dwarf spheroids, but is still hotly debated in the literature \citep[e.g.][]{bin94, gra03, fer06, kormendy09, kor12}.~In order to investigate these possible differences in the formation process of dwarf galaxies in comparison to bright elliptical and spiral galaxies, we plot in Figure~\ref{scalingrelations_reff} the effective
radius ($r_{\rm eff}$,  in units of parsec) as  a function of stellar mass  \mstar$/M_\odot$ and in Figure~\ref{scalingrelations_surfmd} the effective  mass surface density ($\Sigma_{{\rm eff},{\cal  M}_\star}\!=\!{\cal M}_\star/2\pi r_{\rm
eff}^2$, in units of  $M_\odot/{\rm pc}^2$) as a function of  stellar mass \mstar$/M_\odot$ for our NGFS  galaxy sample. We put the measurements  in the larger context of several  other families of stellar systems, covering  a wide range in
galaxy morphology and stellar mass from low-mass LG dSph galaxies with \lmstar~$\!\approx\!3$ to giant elliptical galaxies at \lmstar~$\!\approx\!12$.

For the subsequent discussion, we divide the  $r_{\rm eff}$--\mstar\ and $\Sigma_{{\rm eff},{\cal M}_\star}$--\mstar\ diagrams in two hemispheres with {\it i)} extended  stellar systems ($\log r_{\rm eff}[\rm{pc}]\gtrsim2.0$), such as dwarf
spheroidal galaxies,  dwarf elliptical, and  giant elliptical (gE)  galaxies, and {\it  ii)} compact stellar  systems ($\log r_{\rm eff}[\rm{pc}]\!\lesssim\!2.0$)  with two subsets,  one of which  represents globular clusters  with two-body
relaxation times shorter than a Hubble time (i.e.~objects  below the curve $t_{\rm relax}\!=\!$~Hubble time) and the other subset comprising non-relaxed compact stellar systems, such  as nuclear star clusters (NSCs) and ultra-compact dwarfs
(UCDs).~For a  description of this part  of the parameter  space we refer the  reader to \cite{misgeld}.~We  will not discuss the  compact stellar systems  further in this work,  noting only that  numerous recent studies have  described the
formation scenario of UCDs via tidal  stripping of galaxies with NSCs on relatively short  timescales \citep[e.g.][]{bek03, goe08, tho08, pb13, pfe14, liu, zhang15, pfe16}.~Instead, we  focus in the following on the extended stellar  systems and the larger galaxy evolution context of our NGFS galaxy sample.

The NGFS sample dwarfs cover stellar masses in the range \lmstar~$\!\approx\!5.5\!-\!9.5$ (see also Fig.~\ref{stellarmass}).~The bulk of
the NGFS dwarfs show  \mdens values in the range $\!\sim\!1\!-\!10\,M_\odot$ pc$^{-2}$  with the most massive dwarfs reaching $\!\sim\!100\,M_\odot$ pc$^{-2}$  (see Figs.~\ref{scalingrelations_reff} and \ref{scalingrelations_surfmd}).~Dwarf
galaxies with relatively  low stellar masses (\lmstar~$\!\simeq\!5.5\!-\!8.0$) follow  a scaling relation with $r_{\rm  eff}$ which is positively  inclined with respect to the  lines of equal effective surface  mass density, i.e.~iso-\mdens
lines (see  diagonal lines in Fig.~\ref{scalingrelations_reff}).~Meanwhile, recent studies have  revealed an expansive population  of low-mass dwarfs in  the Local Group, including  dSphs and UFDs that  extend towards even fainter  magnitudes 
\citep{zucker04, zucker06a, zucker06b, willman,bel06, bel07, mcconnachie09, bel10, mcconnachie12, bel14, bec15, drl15, kop15, lae15, hom16}.~Their  size-mass-surface density scaling relations appear to line up seamlessly with our NGFS dwarf sample
towards lower stellar masses and lower surface mass densities.

We  approximate  the scaling  relations  of  our  NGFS  dwarfs with  a  Locally  Weighted  Scatterplot Smoothing  (LOWESS)  fit  \citep[e.g.][]{cel81},  which is  illustrated  as  a  thick  curve in  Figures~\ref{scalingrelations_reff}  and
\ref{scalingrelations_surfmd}.~To connect the LOWESS fit of  NGFS dwarf sample with more massive galaxies we construct two  samples that span the full stellar mass range \lmstar~$\!\approx\!5.5\!-\!12.0$: {\bf  A}) NGFS dwarfs  + Virgo ETGs \citep{fer06} + Carnegie-Irvine E galaxies  \citep{ho11} shown as a brown curve; and {\bf B}) NGFS dwarfs + spiral galaxies from the ATLAS3D survey\footnote{We  point out that we do not include the early-type ATLAS3D
galaxies in sample  {\bf A} due to their  relatively shallow stellar mass limit  ($\sim\!10^{9.5}\,M_\odot$).} \citep{cap11} illustrated as a blue  curve.~While sample {\bf A} assumes  that dwarf galaxies are evolutionary  connected to more
massive early-type  galaxies through their morphology,  sample {\bf B}  joins, somewhat {\it ad  hoc}, the dwarf  galaxy regime with more  massive spirals under  the premise that they  form a dynamical  family in which the  angular momentum
increases with stellar mass \citep[for a discussion, see][]{cap16}.

We  disregard  the class  of  UDGs  for  the rest  of  this  work  \citep[but see][for  a  theoretical  discussion  of this  galaxy  type]{amo16,  ron17},  but note  that  the  scaling  relations in  Figures~\ref{scalingrelations_reff}  and
\ref{scalingrelations_surfmd} show  a few  NGFS objects consistent  with massive dwarf  galaxy candidates  ($7.5\!\lesssim\!\log{\cal M}_\star/M_\odot\!\lesssim\!8.5$) beginning  to encroach upon  the parameter  space occupied by  UDGs, the
existence of which has been recently reported in  massive galaxy groups and clusters \citep[e.g.][]{vandokkum, koda, mihos, mun15, mar16, mer16, tru17, rom17b, jans17, vdb17, shi17, ben17, ven17}.~UDGs show physical sizes  reminiscent of giant galaxies, but \mstar\ values that put
them in the dwarf  galaxy mass regime.~The result is a systematically lower  \mdens\ that detaches UDGs from the scaling relations  shown by the other dwarfs and intermediate stellar-mass  galaxies.~Given their existence and characteristics
one might speculate that these galaxies are dominated by dark-matter, shielding the baryons from the cluster environment  \citep[e.g.][]{mihos,mow17}.

\subsection{Scaling Relation Gradients \label{gradientsection}}
For sample {\bf A},  we find a galaxy size-mass scaling relation  of the form $r_{\rm eff}\!\propto\!{\cal M}_\star^{0.3\pm0.1}$ in  the stellar mass range $5.5\!\lesssim\!\log{\cal M}_\star\!\lesssim\!7.8$, the  gradient of which decreases
abruptly to zero at \lmstar~$\!\approx\!8.0$ (i.e.~$\nabla_{\rm LOWESS}\!=\Delta\log r_{\rm eff}/\Delta\log{\cal M}_\star\!\approx\!0$,  see inset plot in Fig.~\ref{scalingrelations_reff}).~The effective radius remains virtually independent
of  stellar mass  for more  massive galaxies  and remains  roughly constant  at $\sim\!1$\,kpc  (albeit with  a substantial  scatter) up to about  \lmstar~$\!\approx\!9.5$, where  it begins  to gently  increase again  until it  jumps abruptly  at
\lmstar~$\!\approx\!10.5$ to values that are steeper  than the iso-\mdens lines.~The gradient at high stellar masses (\lmstar~$\!\gtrsim\!10.8$) reaches $r_{\rm  eff}\!\propto\!{\cal M}_\star^{0.75\pm0.05}$.~For sample {\bf B}, the gradient
of the size-mass  relation is relatively constant over the  entire mass range (\lmstar~$\!\approx\!5.5\!-\!12$) around $r_{\rm  eff}\!\propto\!{\cal M}_\star^{0.25\pm0.1}$.~Although we are dealing with  incomplete and partly non-overlapping
parameter ranges for the individual datasets depicted  in Figures~\ref{scalingrelations_reff} and \ref{scalingrelations_surfmd}, we observe a smooth transition between the low-mass dwarf  regime over the range of intermediate stellar-mass systems to
the massive galaxies for our self-consistently observed and  analyzed sample of NGFS galaxies. Here, the scaling relations become particularly diagnostic in the intermediate-mass regime ($8.5\!\lesssim\!$~\lmstar~$\!\lesssim\!10.0$), where the association of at least some of the NGFS galaxies with the sample {\bf B} dynamical family cannot be excluded.~Testing the internal dynamics of such galaxies is, therefore, of utmost interest.
~The literature samples are augmenting the information content of the scaling relations for our NGFS data by  extending its parameter space coverage, which
leads us to the following discussion and interpretation of our measurements.

\subsection{Implications for Galaxy Mass Assembly}
Dwarf galaxies in cluster environments are  believed to have been and some are presently being affected by their environment, which is  one of the mechanisms for the creation of the morphology-density  or morphology-distance relation for dwarfs.~Environmental
effects include tidal stripping, ram pressure  stripping, and harassment. Moreover, internal effects may play a role. For  instance, winds have been invoked as one of the factors affecting their  chemical enrichment, and feedback, in general,
may have  led to mass loss.~How the size-stellar mass (i.e.\ $r_{\rm eff}$--\mstar) scaling relation for extended stellar  systems can be  interpreted in the context  of the hierarchical galaxy  formation picture has been  extensively discussed in the  recent literature
\citep[for a recent review see][and references therein]{cap16}.

Based on the present NGFS data and  the comparison with other stellar systems from the literature, the following discussion  aims at highlighting the implications of our findings for galaxy mass assembly  over a stellar mass range of $5
\lesssim \log$ \mstar$\lesssim  12$. We want to  emphasize that this discussion  does not necessarily imply an  evolutionary sequence in the  sense that more massive stellar  systems arise from less  massive ones observed today. It is  merely an empirical
observation based on the  derived gradients from Sect.\ \ref{gradientsection} on how mass  assembly occurs in these galaxies at various stellar  masses.

First, we note that mass assembly  along the iso-\mdens lines implies virtually density-invariant growth, indicating that to  first order, mass assembly occurs homologuously along these lines, i.e.\ as  galaxies grow in size and stellar
mass they retain their surface brightness distribution and, hence, the effective stellar mass surface density.

From Figure~\ref{scalingrelations_reff}  we find that as  early-type systems (sample {\bf  A}) grow in stellar  mass, the assembly of baryonic  mass must occur in  various phases and under the influence of  different mechanisms, which depend  on the total
stellar mass of  the system, for the observed relations to emerge. For dwarf galaxies  this mass  assembly process  occurs biased to regions inside the  half-light/mass radius since the systems  grow 
disproportionally more dense within  $r_{\rm eff}$ as stellar  mass increases, i.e.~their  $r_{\rm eff}$--\mstar relation is   flatter with  respect to the  iso-\mdens lines.~This trend prevails  until \mstar\
approaches \lmstar~$\!\approx\!8.0$, where  the $r_{\rm eff}$--\mstar\ relation becomes flat  with significant scatter. The constancy of  $r_{\rm eff}$ in this respective mass  range had already been pointed out  by \citet{smcast08} and
\citet{misgeld08}  and implies  that stellar mass  is  being added  both inside  and  outside the  galaxy half-light/stellar-mass  radii  so that  galaxy stellar  mass  growth occurs  without experiencing  changes  in galaxy  sizes, i.e.\  isometric.
Consequently,  this  mode of  stellar-mass  growth is  accompanied  by  packing increasingly  more  stellar  mass within  the  same  galaxy dimensions.~This process may be triggered by ram-pressure in dense galaxy cluster environments which may  
temporarily induce enhanced star formation efficiency \citep[e.g.][]{geha12, wet13}.~The isometric stellar mass growth  stops  when the  galaxies  reach  an apparent  maximum  stellar  mass density  of
\mdens~$\!\approx\!10^4\,M_\odot$\,pc$^{-2}$ at  around \lmstar~$\!\approx\!10.5$.~Beyond  this point  towards higher \mstar\  values, galaxy  stellar mass  accumulation happens predominantly  outside their  half-light/mass radii,  which is
consistent with  their $r_{\rm eff}$--\mstar\  relations being steeper than  the iso-\mdens\ lines.~Similar  findings have been  put forward by studies  of the redshift  evolution of the  $r_{\rm eff}$--\mstar\ relation of  massive galaxies
\citep[e.g.][]{hue13, vdokk15}.

This  observed transition  from  centrally-dominated towards  halo-dominated  stellar  mass accumulation  goes  hand in  hand  with the  transition  from cusp  to  core-dominated central  surface  brightness  profiles of  galaxies  more massive  than
\lmstar~$\!\approx\!9.0$ revealed by HST-based studies \citep[see e.g.][]{fer06, cot07}.~The  homogeneous morphological analysis of our NGFS galaxy dataset, with a much broader stellar mass  coverage down to \lmstar~$\!\approx\!5.5$, shows the consistency
of this central  morphological transition with global S\'ersic model  parameters.~While single-component S\'ersic fits represent the  majority of our NGFS dwarfs very  well (cf.\ Fig.~\ref{galaxymodeling}), the surface-brightness profile  fits of the
most massive NGFS galaxies  (\lmstar~$\!\gtrsim\!9.0$) always show residuals, indicative of substructure  in single-component fits, requiring up to two  additional significant S\'ersic components for the brightest  systems.~This corroborates the idea
that the mass-assembly process of  dSph and gE galaxies sees the predominance of different  mechanisms in action, with the former being mainly assembled through  dissipative processes such as wet mergers and gas infall  that build the central stellar
mass and lead  to surface-brightness profiles well represented by  single-component S\'ersic models \citep{fer06}.~The gE galaxies,  on the other hand, form predominantly  via non-dissipative stellar mass accumulation (i.e.\  dry galaxy mergers) that
preserve the phase-space configuration of  the infalling material, which results in more complex  surface-brightness profiles that are indicative of halo mass  accumulation. This bimodal galaxy growth is partly driven  by the increasing importance of
nuclear feedback processes with increasing total galaxy mass and has been extensively discussed in the  recent literature \citep[e.g.][]{cro06, dek06, fab12, kor13}.~Our study delivers a large dwarf galaxy sample, which can
be used to constrain the details of quenching mechanisms in today's high-resolution numerical simulations \citep[e.g.][]{gen17, bah17}.

Future analyses using the NGFS galaxy sample from the entire survey footprint with an augmented SED filter coverage will  allow us to conduct more extended in-depth studies as a function of the cluster-centric radius, and search for local anomalies
that correlate with particular families of galaxies in the size-mass-surface density parameter space.

\section{Summary and Conclusion}
This paper is part  of a series utilizing newly taken deep $u'g'i'$  photometry from the Next Generation Fornax Survey (NGFS)  obtained with the Dark Energy Camera mounted at  the prime focus of the Blanco 4-m telescope  at the Cerro Tololo
Interamerican Observatory  in Chile.~In the  present work we analyze  the faint and  bright galaxy population  in the core of  the Fornax galaxy  cluster via photometric  and structural parameters, measured  using the {\sc  galfit} software
package.~In total,  258 dwarf  galaxy candidates have  been identified  by visual inspection  of deep  RGB image stacks,  with galaxy  spheroid components reconstructed  by single-  and multi-component S\'ersic  models.~The current  analysis is based on the central $\lesssim 0.25R_{\rm vir}$  tile of the NGFS survey  footprint and will be complemented  with the remaining survey data  in future
work.~In the following we summarize our main results:

\begin{itemize}

\item The morphological analysis of  the dwarf galaxies shows that their spheroid surface-brightness  profiles are well represented by single-component S\'ersic models with  average S\'ersic indices in the well-constrained range of $\langle n\rangle_{u',g',i'}\!=\!(0.78\!-\!0.83)\pm0.02$, and average effective radii in the range of $\langle r_e\rangle_{u',g',i'}\!=\!(0.67\!-\!0.70)\pm0.02$\,kpc.

\item We find  that 75/258 ($\sim$29\%) of the  dwarf galaxy candidates are nucleated  based on visual inspection.~Comparing galaxy morphologies with total galaxy magnitudes, we find that the nucleation fraction (\fnuc) is a  strong function of galaxy luminosity and \mstar. \fnuc\ shows a strong tendency towards higher fractions in bins containing
the brightest galaxies, and declines to $0\%$ for faint galaxies ($M_{\rm g'}\!\gtrsim\!-9$).

%\item We  estimate a  Fornax galaxy  luminosity function faint-end  slope of  $\alpha=-1.45\pm 0.15$  for galaxy luminosities  in the range  $-12\!\gtrsim\!M_{g'}\!\gtrsim\!-15$\,mag, corresponding  to $7.3\lesssim  \log{\cal M}_{\star}/M_\odot
%\lesssim\!8.6$.~This value is comparable with other galaxy groups and clusters and is much shallower than the expected halo mass function slope from $\Lambda$CDM cosmology.

\item In multi-filter color-magnitude diagrams, the Fornax dwarf galaxy population follows a typical cluster red sequence with brighter galaxies generally exhibiting redder colors.~The red sequence slope matches that observed for the dwarf
galaxy population in the core of the Virgo cluster.~We observe a significant flattening of the red sequence at the faintest galaxy magnitudes, similar to that observed in Virgo.~However, the flattening occurs at bluer colors.~Because of the color-luminosity relation and the brighter average magnitudes of nucleated dwarfs, the colors of nucleated spheroids are, on average, redder than those of non-nucleated spheroids.

\item Bruzual-Charlot SSP models computed for  $u'g'i'$ bands indicate that the investigated Fornax dwarf galaxy population  is, on average, older than $\sim\!1\,$Gyr if the galaxy metallicities  are solar-type.~Assuming that our dwarf
sample exhibits much  lower metallicites based on  the galaxy mass-metallicity relation, the  average dwarf in our  NGFS sample is likely older  than $\sim\!5$ Gyr.~Dwarf  galaxy colors are also consistent  with a large metallicity  spread if the
galaxies are uniformly older than $\sim\!10$\,Gyr.~Spectroscopic information or near-IR photometry is needed to break this age-metallicity degeneracy.

\item The Fornax  dwarf galaxy population is  consistent with known scaling relations  between half-light/stellar-mass radius, stellar  mass, and stellar mass  density for dwarf galaxies, i.e.\  they follow increasing relations  of surface-brightness and/or
stellar mass surface density with increasing luminosity and/or stellar mass.~For the brightest dwarf galaxies this  correlation flattens and reflects the transition towards bright ellipticals showing a presumably more complex formation scenario.

\item We find that over the sampled stellar mass range several distinct mechanisms of galaxy mass  assembly can be identified: {\it i)} dwarf galaxies assemble mass
inside the half-mass  radius up to \lmstar~$\!\approx\!8.0$,  {\it ii)} isometric mass assembly  in the range $8.0\lesssim \log{\cal  M}_{\star}/M_\odot \lesssim10.5$, and {\it  iii)} massive galaxies assemble stellar  mass predominantly in
their halos at \lmstar~$\!\approx\!10.5$ and above.
\end{itemize}

\acknowledgments

This project is supported by FONDECYT Postdoctoral  Fellowship Project No.~3130750, FONDECYT Regular Project No.~1161817 and the BASAL Center for Astrophysics and  Associated Technologies (PFB-06).~P.E.~acknowledges support from the Chinese
Academy of Sciences (CAS) through CAS-CONICYT Postdoctoral  Fellowship CAS150023 administered by the CAS South America Center for Astronomy (CASSACA) in Santiago,  Chile.~M.A.T.~acknowledges the financial support through an excellence grant
from the Vicerrector\'ia de Investigaci\'on  and the Institute of Astrophysics Graduate School Fund at  Pontificia Universidad Cat\'olica de Chile and the European Southern  Observatory Graduate Student Fellowship program.~G.G.~acknowledges
support from FONDECYT Regular Project No.~1120195. Y.O.-B.\ acknowledges financial support through CONICYT-Chile (grant CONICYT-PCHA/Doctorado Nacional/2014-21140651). A.L.\ and M.P. acknowledge ECOS-Sud/CONICYT project C15U02.

This project used data  obtained with the Dark Energy Camera (DECam), which  was constructed by the Dark Energy Survey  (DES) collaboration. Funding for the DES Projects has  been provided by the DOE and NSF (USA),  MISE (Spain), STFC (UK),
HEFCE (UK), NCSA (UIUC), KICP (U.\ Chicago), CCAPP (Ohio  State), MIFPA (Texas A\&M), CNPQ, FAPERJ, FINEP (Brazil), MINECO (Spain), DFG (Germany) and the collaborating institutions in the  Dark Energy Survey, which are Argonne Lab, UC Santa
Cruz, University of  Cambridge, CIEMAT-Madrid, University of Chicago,  University College London, DES-Brazil Consortium,  University of Edinburgh, ETH Z{\"u}rich, Fermilab,  University of Illinois, ICE (IEEC-CSIC),  IFAE Barcelona, Lawrence
Berkeley Lab, LMU M{\"u}nchen and the associated Excellence  Cluster Universe, University of Michigan, NOAO, University of Nottingham, Ohio State University, University of  Pennsylvania, University of Portsmouth, SLAC National Lab, Stanford
University, University of Sussex, and Texas A\&M University. \\

This research has  made use of the NASA Astrophysics  Data System Bibliographic Services, the  NASA Extragalactic Database, the SIMBAD database,  operated at CDS, Strasbourg, France  \citep{wen00}. This research has made use  of "Aladin Sky
Atlas" \citep{bon00} developed at  CDS, Strasbourg Observatory, France. Software used in  the analysis includes the data analysis algorithm {\sc  galfit} \citep[v3.0.5;][]{pen10}, the astronomical data reduction package  {\sc IRAF} which is
distributed  by  the   National  Optical  Astronomy  Observatories  (NOAO)  as   well  as  the  {\sc  Python/NumPy}   v.1.11.2  and  {\sc  Python/Scipy}  v0.17.0  \citep[][\url{http://www.scipy.org/}]{jon01,   van11},  {\sc  Python/astropy}
\citep[v1.2.1;][\url{http://www.astropy.org/}]{ast13}, {\sc Python/matplotlib} \citep[v2.0.0;][\url{http://matplotlib.org/}]{hun07}, {\sc Python/scikit-learn} \citep[v0.17.1;][\url{http://scikit-learn.org/stable/}]{ped12} packages.\\

{\it Facilities:} \facility{CTIO (4m Blanco/DECam)}.

\appendix
\section{Dwarf Galaxy Sample Photometric Data}
\begin{deluxetable*}{cccccccccccc}
%\tablewidth{400pt}
\tablecaption{Photometric properties and stellar masses of the dwarf galaxy sample\label{tab:dwarfphot}}
%-------------------------------------------------------------------------------------------------------------------------------------------------------------------------------------------------------------------------------------------------------------------------------------------------------------------------------------------------------------------------------------------------------------------
\tablehead{                                                                                                                                                                                                                                                                                                                               
\colhead{ID}        &  \colhead{Reference\tablenotemark{a}}          & \colhead{$\alpha$}              &  \colhead{$\delta$}                           &  \colhead{$i'$}  	&  \colhead{$g'$}  	&  \colhead{$u'$}  	&  \colhead{$(g'-i')_{0}$}  &  \colhead{$(u'-g')_{0}$} &  \colhead{$(u'-i')_{0}$} &  \colhead{$M_g$\tablenotemark{b}}  & \colhead{$\log {\cal M}_\star$} \\
\colhead{}		    &  \colhead{}								     & \colhead{(J2000)}		       &  \colhead{(J2000)}	                           &  \colhead{(mag)}	&  \colhead{(mag)}	&  \colhead{(mag)}	&  \colhead{(mag)}	        &  \colhead{(mag)}		   &  \colhead{(mag)}	      &  \colhead{(mag)}				   & \colhead{($M_\odot$)}
}    
%-------------------------------------------------------------------------------------------------------------------------------------------------------------------------------------------------------------------------------------------------------------------------------------------------------------------------------------------------------------------------------------------------------------------
\startdata                                                                                                                                                                                                                                                                                                                                                                        
NGFS033309-352349   &    FCC114                                      &    03$^{h}$33$^{m}$08$\fs$63    &     $-$35$\arcdeg$23$\arcmin$49$\farcs$01     &    18.67           &  19.57            &   20.28           &  0.90                     &  0.72                    &  1.62                    &  $-$11.94                          &   7.11                          \\
NGFS033311-353956   &    \nodata                                     &    03$^{h}$33$^{m}$10$\fs$93    &     $-$35$\arcdeg$39$\arcmin$56$\farcs$16     &    20.05           &  21.62            &   \nodata         &  1.57                     &  \nodata                 &  \nodata                 &   $-$9.89                          &   7.05                          \\
NGFS033348-355010   &    FCC125                                      &    03$^{h}$33$^{m}$48$\fs$42    &     $-$35$\arcdeg$50$\arcmin$09$\farcs$66     &    17.64           &  18.50            &   19.64           &  0.86                     &  1.15                    &  2.01                    &  $-$13.01                          &   7.62                          \\
NGFS033350-355706   &    \nodata                                     &    03$^{h}$33$^{m}$49$\fs$87    &     $-$35$\arcdeg$57$\arcmin$06$\farcs$17     &    20.91           &  21.57            &   \nodata         &  0.66                     &  \nodata                 &  \nodata                 &   $-$9.94                          &   6.24                          \\
NGFS033400-354533   &    \nodata                                     &    03$^{h}$33$^{m}$59$\fs$75    &     $-$35$\arcdeg$45$\arcmin$33$\farcs$26     &    18.30           &  19.52            &   20.05           &  1.22                     &  0.54                    &  1.76                    &  $-$11.99                          &   7.27                          \\
NGFS033406-351638   &    FCC127                                      &    03$^{h}$34$^{m}$05$\fs$97    &     $-$35$\arcdeg$16$\arcmin$38$\farcs$25     &    18.61           &  19.39            &   20.30           &  0.78                     &  0.92                    &  1.70                    &  $-$12.12                          &   7.17                          \\
NGFS033407-352838   &    \nodata                                     &    03$^{h}$34$^{m}$06$\fs$97    &     $-$35$\arcdeg$28$\arcmin$37$\farcs$99     &    20.63           &  22.47            &   \nodata         &  1.84                     &  \nodata                 &  \nodata                 &   $-$9.04                          &   6.96                          \\
NGFS033409-353100   &    FCC130                                      &    03$^{h}$34$^{m}$09$\fs$18    &     $-$35$\arcdeg$30$\arcmin$59$\farcs$50     &    17.38           &  18.10            &   \nodata         &  0.72                     &  \nodata                 &  \nodata                 &  $-$13.41                          &   7.68                          \\
NGFS033412-351343   &    FCC131                                      &    03$^{h}$34$^{m}$12$\fs$13    &     $-$35$\arcdeg$13$\arcmin$43$\farcs$28     &    19.15           &  19.92            &   20.50           &  0.77                     &  0.59                    &  1.36                    &  $-$11.59                          &   6.88                          \\
NGFS033414-354910   &    \nodata                                     &    03$^{h}$34$^{m}$14$\fs$29    &     $-$35$\arcdeg$49$\arcmin$10$\farcs$05     &    19.79           &  20.47            &   \nodata         &  0.68                     &  \nodata                 &  \nodata                 &  $-$11.04                          &   6.70                          \\
NGFS033423-355042   &    \nodata                                     &    03$^{h}$34$^{m}$23$\fs$34    &     $-$35$\arcdeg$50$\arcmin$41$\farcs$89     &    19.18           &  20.51            &   \nodata         &  1.33                     &  \nodata                 &  \nodata                 &  $-$11.00                          &   7.28                          \\
NGFS033427-350621   &    \nodata                                     &    03$^{h}$34$^{m}$27$\fs$18    &     $-$35$\arcdeg$06$\arcmin$20$\farcs$76     &    20.55           &  21.55            &   \nodata         &  1.00                     &  \nodata                 &  \nodata                 &   $-$9.96                          &   6.56                          \\
NGFS033433-350236   &    \nodata                                     &    03$^{h}$34$^{m}$32$\fs$55    &     $-$35$\arcdeg$02$\arcmin$35$\farcs$50     &    20.38           &  20.62            &   \nodata         &  0.24                     &  \nodata                 &  \nodata                 &  $-$10.89                          &   6.23                          \\
NGFS033436-360315   &    \nodata                                     &    03$^{h}$34$^{m}$36$\fs$38    &     $-$36$\arcdeg$03$\arcmin$15$\farcs$27     &    20.56           &  21.61            &   \nodata         &  1.05                     &  \nodata                 &  \nodata                 &   $-$9.90                          &   6.58                          \\
NGFS033443-353115   &    \nodata                                     &    03$^{h}$34$^{m}$43$\fs$39    &     $-$35$\arcdeg$31$\arcmin$15$\farcs$07     &    20.45           &  21.04            &   \nodata         &  0.59                     &  \nodata                 &  \nodata                 &  $-$10.47                          &   6.39                          \\
NGFS033446-345334   &    \nodata                                     &    03$^{h}$34$^{m}$46$\fs$06    &     $-$34$\arcdeg$53$\arcmin$33$\farcs$56     &    20.52           &  21.29            &   \nodata         &  0.77                     &  \nodata                 &  \nodata                 &  $-$10.22                          &   6.45                          \\
NGFS033456-351127   &    FCC140                                      &    03$^{h}$34$^{m}$56$\fs$41    &     $-$35$\arcdeg$11$\arcmin$27$\farcs$42     &    17.61           &  18.25            &   19.37           &  0.64                     &  1.12                    &  1.76                    &  $-$13.26                          &   7.59                          \\
NGFS033458-351324   &    WFLSB6-2                                    &    03$^{h}$34$^{m}$57$\fs$56    &     $-$35$\arcdeg$13$\arcmin$24$\farcs$42     &    18.72           &  19.72            &   20.45           &  1.00                     &  0.74                    &  1.74                    &  $-$11.79                          &   7.11                          \\
NGFS033458-350235   &    FCC142                                      &    03$^{h}$34$^{m}$58$\fs$19    &     $-$35$\arcdeg$02$\arcmin$34$\farcs$80     &    17.34           &  18.37            &   19.45           &  1.03                     &  1.08                    &  2.11                    &  $-$13.14                          &   7.74                          \\
NGFS033500-351920   &    FCC144                                      &    03$^{h}$35$^{m}$00$\fs$20    &     $-$35$\arcdeg$19$\arcmin$20$\farcs$26     &    18.24           &  19.11            &   19.83           &  0.87                     &  0.73                    &  1.60                    &  $-$12.40                          &   7.28                          \\\vspace{-0.25cm}
\enddata
\tablenotetext{a}{Reference to galaxies listed in the FCC catalogue \citep{ferguson89} and in \citet{mie07}.}
\tablenotetext{b}{Assuming a distance modulus of $(m\!-\!M)_0\!=\!31.51$ mag \citep{bla09}.}
\tablecomments{Table\,\ref{tab:dwarfphot} is published in its entirety in the electronic edition of the {\it Astrophysical Journal}. A portion is shown here for guidance regarding its form and content.}
\end{deluxetable*}

\begin{deluxetable*}{ccc|cccc|cccc|cccc}
%\tablewidth{420pt}
\tablecaption{Structural properties of the dwarf sample\label{tab:dwarfmorph}}
%-------------------------------------------------------------------------------------------------------------------------------------------------------------------------------------------------------------------------------------------------------------------------------------------------------------------------------------------------------------------------------------------------------------------
\tablehead{ \colhead{\multirow{2}{*}{ID}}    &    \colhead{\multirow{2}{*}{Reference}}   &  \colhead{\multirow{2}{*}{Type\tablenotemark{a}}}   &                                         \multicolumn{4}{c}{$i'$}                                                                                    &                                             \multicolumn{4}{c}{$g'$}                                                                                &                                        \multicolumn{4}{c}{$u'$}                                                          \\
                                             &                                           &                                                     & \colhead{${r_e}$\tablenotemark{b}}   &  \colhead{$n$\tablenotemark{c}}  &  \colhead{$\epsilon$\tablenotemark{d}}  &  \colhead{PA\tablenotemark{e}}  &  \colhead{${r_e}$\tablenotemark{b}}  &  \colhead{$n$\tablenotemark{c}}  &  \colhead{$\epsilon$\tablenotemark{d}}  &  \colhead{PA\tablenotemark{e}}  & \colhead{${r_e}$\tablenotemark{b}}  &    \colhead{$n$\tablenotemark{c}}  &  \colhead{$\epsilon$\tablenotemark{d}}  &  \colhead{PA\tablenotemark{e}} }                   
%-------------------------------------------------------------------------------------------------------------------------------------------------------------------------------------------------------------------------------------------------------------------------------------------------------------------------------------------------------------------------------------------------------------------
\startdata                                                                                                                                                                                                                                                                                                                                                                        
NGFS033309-352349                            &    FCC114                                 &    $\medcircle$                                     &                  0.699               &                 0.81             &                       0.43              &          98                     &           0.642                      &           0.72                   &        0.47                             &      97                         &        0.584                        &         0.64                      &             0.46                         &      99                  \\ 
NGFS033311-353956                            &    \nodata                                &    $\medcircle$                                     &                  0.883               &                 0.63             &                       0.24              &         125                     &           0.757                      &           0.23                   &        0.50                             &     106                         &        \nodata                      &         \nodata                   &             \nodata                      &      \nodata             \\      
NGFS033348-355010                            &    FCC125                                 &    $\medcircle$                                     &                  0.896               &                 0.71             &                       0.15              &         141                     &           0.866                      &           0.65                   &        0.14                             &     139                         &        0.766                        &         0.50                      &             0.13                         &      150                 \\  
NGFS033350-355706                            &    \nodata                                &    $\medcircle$                                     &                  0.435               &                 0.45             &                       0.39              &         114                     &           0.503                      &           0.50                   &        0.43                             &     136                         &        \nodata                      &         \nodata                   &             \nodata                      &      \nodata             \\      
NGFS033400-354533                            &    \nodata                                &    $\medcircle$                                     &                  0.563               &                 0.87             &                       0.24              &          21                     &           0.421                      &           0.63                   &        0.17                             &      30                         &        0.622                        &         1.13                      &             0.30                         &      29                  \\ 
NGFS033406-351638                            &    FCC127                                 &    $\medcircle$                                     &                  0.619               &                 0.82             &                       0.23              &          72                     &           0.602                      &           0.82                   &        0.22                             &      74                         &        0.526                        &         0.76                      &             0.24                         &      73                  \\ 
NGFS033407-352838                            &    \nodata                                &    $\medcircle$                                     &                  0.623               &                 0.82             &                       0.13              &         173                     &           0.359                      &           0.30                   &        0.16                             &     118                         &        \nodata                      &         \nodata                   &             \nodata                      &      \nodata             \\      
NGFS033409-353100                            &    FCC130                                 &    $\medcircle$                                     &                  1.819               &                 0.80             &                       0.17              &         135                     &           2.008                      &           0.91                   &        0.18                             &     130                         &        \nodata                      &         \nodata                   &             \nodata                      &      \nodata             \\      
NGFS033412-351343                            &    FCC131                                 &    $\odot$                                          &                  0.526               &                 0.75             &                       0.44              &         127                     &           0.521                      &           0.91                   &        0.47                             &     128                         &        0.475                        &         0.72                      &             0.34                         &      109                 \\  
NGFS033414-354910                            &    \nodata                                &    $\medcircle$                                     &                  0.634               &                 0.39             &                       0.17              &           2                     &           0.715                      &           0.57                   &        0.25                             &     179                         &        \nodata                      &         \nodata                   &             \nodata                      &      \nodata             \\      
NGFS033423-355042                            &    \nodata                                &    $\medcircle$                                     &                  0.848               &                 0.73             &                       0.21              &          98                     &           0.673                      &           0.41                   &        0.15                             &      69                         &        \nodata                      &         \nodata                   &             \nodata                      &      \nodata             \\      
NGFS033427-350621                            &    \nodata                                &    $\medcircle$                                     &                  0.466               &                 0.95             &                       0.42              &          26                     &           0.356                      &           0.68                   &        0.32                             &      33                         &        \nodata                      &         \nodata                   &             \nodata                      &      \nodata             \\      
NGFS033433-350236                            &    \nodata                                &    $\medcircle$                                     &                  0.499               &                 0.78             &                       0.36              &         100                     &           0.657                      &           0.69                   &        0.39                             &      99                         &        \nodata                      &         \nodata                   &             \nodata                      &      \nodata             \\      
NGFS033436-360315                            &    \nodata                                &    $\medcircle$                                     &                  0.573               &                 0.62             &                       0.51              &         168                     &           0.467                      &           0.50                   &        0.36                             &     179                         &        \nodata                      &         \nodata                   &             \nodata                      &      \nodata             \\      
NGFS033443-353115                            &    \nodata                                &    $\medcircle$                                     &                  0.377               &                 0.74             &                       0.09              &          99                     &           0.400                      &           0.73                   &        0.11                             &      89                         &        \nodata                      &         \nodata                   &             \nodata                      &      \nodata             \\      
NGFS033446-345334                            &    \nodata                                &    $\odot$                                          &                  0.440               &                 0.65             &                       0.29              &         140                     &           0.410                      &           0.67                   &        0.22                             &      98                         &        \nodata                      &         \nodata                   &             \nodata                      &      \nodata             \\      
NGFS033456-351127                            &    FCC140                                 &    $\odot$                                          &                  0.898               &                 0.81             &                       0.46              &          95                     &           0.968                      &           0.89                   &        0.47                             &      94                         &        0.795                        &         0.63                      &             0.47                         &      94                  \\ 
NGFS033458-351324                            &    WFLSB6-2                               &    $\medcircle$                                     &                  0.559               &                 1.00             &                       0.29              &          41                     &           0.508                      &           0.80                   &        0.31                             &      50                         &        0.486                        &         0.65                      &             0.30                         &      54                  \\ 
NGFS033458-350235                            &    FCC142                                 &    $\odot$                                          &                  1.011               &                 1.34             &                       0.06              &         134                     &           0.847                      &           1.02                   &        0.06                             &     149                         &        0.708                        &         0.92                      &             0.11                         &      16                  \\ 
NGFS033500-351920                            &    FCC144                                 &    $\medcircle$                                     &                  0.597               &                 0.81             &                       0.23              &          56                     &           0.592                      &           0.77                   &        0.25                             &      56                         &        0.523                        &         0.82                      &             0.23                         &      55                  \\ 
\enddata
\tablenotetext{a}{Morphological galaxy type classification: $\odot$=nucleated, $\medcircle$=non-nucleated dwarf galaxy.}
\tablenotetext{b}{Effective radii given in kpc.}
\tablenotetext{c}{S\'{e}rsic index \citep{ser68, cao93}.}
\tablenotetext{d}{Ellipticity $\epsilon=1-b/a$.}
\tablenotetext{e}{Position angle in $\arcdeg$ from North towards East.}
\tablecomments{Table\,\ref{tab:dwarfmorph} is published in its entirety in the electronic edition of the {\it Astrophysical Journal}. A portion is shown here for guidance regarding its form and content.}
\end{deluxetable*}

\end{document}